\batchmode
\makeatletter
\def\input@path{{/Users/sheepawe/Library/CloudStorage/Dropbox/Matrix-Kuramoto-3rd/arxiv/}}
\makeatother
\documentclass[twoside,english]{elsarticle}
\usepackage[T1]{fontenc}
\usepackage[latin9]{inputenc}
\usepackage{geometry}
\usepackage{color}
\usepackage{amsmath}
\usepackage{amsthm}
\usepackage{amssymb}
\usepackage{graphicx}

\makeatletter

\providecommand{\tabularnewline}{\\}

\theoremstyle{plain}
\newtheorem{thm}{\protect\theoremname}
\theoremstyle{plain}
\newtheorem{lem}[thm]{\protect\lemmaname}

\@ifundefined{date}{}{\date{}}

\usepackage{tikz}
\usetikzlibrary{arrows}
\usetikzlibrary{shadows}
\tikzset{
  every overlay node/.style={
    draw=white,anchor=north west,
  },
}
\usetikzlibrary{positioning, decorations.pathmorphing}

\renewcommand{\fnum@figure}{FIG.~\thefigure}

\makeatother

\usepackage{babel}
\providecommand{\lemmaname}{Lemma}
\providecommand{\theoremname}{Theorem}

\begin{document}

\begin{frontmatter}{}

\title{Binary system modes of matrix-coupled multidimensional Kuramoto oscillators
}

\author[rvt]{Chongzhi Wang}

\author[rvt]{Haibin Shao}

\author[rvt]{Dewei Li\corref{cor1}}

\ead{dwli@sjtu.edu.cn}

\cortext[cor1]{Principal corresponding author}

\address[rvt]{Department of Automation, Shanghai Jiao Tong University, Key Laboratory
of System Control and Information Processing, Ministry of Education
of China, Shanghai Engineering Research Center of Intelligent Control
and Management, Shanghai, 200240, China}

\address{}
\begin{abstract}
Synchronization and desynchronization are the two ends on the spectrum
of emergent phenomena that somehow often coexist in biological, neuronal,
and physical networks. However, previous studies essentially regard
their coexistence as a partition of the network units: those that
are in relative synchrony and those that are not. In real-world systems,
desynchrony bears subtler divisions because the interacting units
are high-dimensional, like fish schooling with varying speeds in a
circle, synchrony and desynchrony may occur to the different \emph{dimensions}
of the population. In this work, we show with an ensemble of multidimensional
Kuramoto oscillators, that this property is generalizable to arbitrary
dimensions: for a $d$ dimensional population, there exist $2^{d}$
system modes where each dimension is either synchronized or desynchronized,
represented by a set of almost binary order parameters. Such phenomena
are induced by a matrix coupling mechanism that goes beyond the conventional
scalar-valued coupling by capturing the inter-dimensional dependence
amongst multidimensional individuals, which arises naturally from
physical, sociological and engineering systems. As verified by our
theory, the property of the coupling matrix thoroughly affects the
emergent system modes and the phase transitions toward them. By numerically
demonstrating that these system modes are interchangeable through
matrix manipulation, we also observe explosive synchronization/desynchronization
that is induced without the conditions that are previously deemed
essential. Our discovery provides theoretical analogy to the cerebral
activity where the resting state and the activated state coexist unihemispherically,
it also evokes a new possibility of information storage in oscillatory
neural networks. 
\end{abstract}

\end{frontmatter}{}

\newpage{}

\section*{Introduction}

Emergent macroscopic behavior from local interactions of a large ensemble
has been both an observation captured in biological, physical, or
sociological systems \citep{Boccaletti2018}, and a long-standing
topic of engineering significance \citep{jadbabaie2003coordination,Ren2005}.
Such emergent phenomena are not limited to synchronization and oscillation
in brain science \citep{fell2011role,varela2001brainweb,buzsaki2004neuronal},
divergence or consensus in opinion dynamics \citep{friedkin2016network,Ye2020ContinuoustimeOD},
and clustering in self-sustained bacterial turbulence \citep{wensink2012meso,dunkel2013fluid}.
It is believed that when the modeling of the locality leads to a close
analogy to the reality on macroscopic scale, it also provides key
factors in understanding the reason that such reality occurs. Usually,
such real-world systems take the abstraction of a networked dynamic
system where each unit is associated to a time-dependent variable,
be it scalar-valued or vectorized, and an underlying topology specifies
the scope and strength of interactions that affect the unit. Whatever
form the variables have adopted, it is accepted that such interactions
can be regulated by a scalar-valued parameter termed the coupling
strength.

Yet already, from problems of engineering and sociological backgrounds,
the matrix coupling mechanism arises as a necessary component in characterizing
inter-dimensional dependence amongst multidimensional individuals.
There is direct evidence, as ref. \citep{tuna2019synchronization}
that sets up arrays of coupled LC oscillators and coupled three-link
pendulums, where the discrepancy of each pair of units need to be
weighed by a positive (semi-)definite matrix that corresponds to either
the dissipative coupling or the restorative coupling. Ref. \citep{tuna2019synchronization}
then examined the parameter space for conditions that guarantee synchronized
oscillations. A more intuitive application is found in the context
of opinion dynamics \citep{Ye2020ContinuoustimeOD}, where the exchanges
of opinions between individuals are modified by a constant matrix
representing the logical interdependence of different topics, which
is decisive in the ultimate distribution of opinions amongst the population
and the conclusion to be drawn. Other scenarios of engineering relevance
that utilize the matrix coupling mechanism include bearing-based formation
control \citep{zhao2016localizability}, graph effective resistance
implemented on distributed control and estimation \citep{barooah2008estimation},
and consensus in multi-agent systems \citep{TRINH2018415,Pan2019,wang2022characterizing}.

Such endeavors however, have rarely been introduced to the biology
community where much research interest, from animal flocking, cortical
activity, to synaptic plasticity, involves processes that necessarily
happen on several dimensions. Examples are fish shoaling and schooling
\citep{Partridge1982,Parrish2002}, the interplay of excitatory and
inhibitory populations between different cortical regions \citep{honey2007network,isaacson2011inhibition,Sadilek2015},
and the electrical and chemical synaptic transmissions that are prevalent
in the nervous system \citep{Purves2012}. Notably, an interesting
property that sometimes accompanies these processes is that synchrony
and desynchrony coexsit in a dimension-by-dimension fashion, like
when fish schooling in circles, the individuals may not have any speed
on the $z$ direction which indicates synchronization, but possess
contrasting velocities on the $x\text{-}y$ plane that vary from those
on the inner circle to those on the outer circle. In this research,
we attempt to bridge the above gap by studying the matrix coupling
mechanism on a multidimensional population that largely retains the
setup of the standard Kuramoto model, due to which the mentioned property
is generalizeable to arbitrary dimensions. Proposed as a mathematically-tractable
model on the all-to-all sinusoidally coupled one-dimensional oscillators
\citep{kuramoto2003chemical,strogatz2000kuramoto,van1993lyapunov},
the Kuramoto model serves as a paradigm in describing synchronization
phenomena in biological systems such as collective animal behavior
\citep{buck1968mechanism,acebron2005kuramoto}, circadian rhythms
\citep{childs2008stability,antonsen2008external}, and activity of
neuronal networks \citep{hoppensteadt1997weakly}. Along with its
many generalizations \citep{tanaka2014solvable,zhu2013synchronization,chandra2019continuous,zhang2015explosive},
it is also used to capture physical systems as the Josephson junction
arrays \citep{watanabe1994constants} and power-grid networks \citep{dorfler2012synchronization,dorfler2013synchronization}.

With the matrix-coupled multidimensional Kuramoto model we proposed,
it is soon revealed that due to the principle of matrix multiplication,
any component of an oscillator will form pairwise interactions with
components of other oscillators on the same dimension, but is implicitly
involved in a three-way interaction in the case of components from
any other dimensions. As a result, the proposed model can be viewed
as a network embedded in a simplicial complex of links and triangles,
to connect with a broader context. The simplicial complex has been
under intensive investigation in conjunction with the Kuramoto model
\citep{Millan2019ExplosiveHK,Skardal2019AbruptDA,Skardal2020HigherOI,Skardal2022MultistabilityIC},
in part, because how cliques, or higher-order interactions were consistently
identified in the functional or structural networks of the brain \citep{Reimann2017,Giusti2016TwosCT,Sizemore2016CliquesAC}.
We mention that, although an implicit higher-order interaction is
considered for components of different dimensions, its mathematical
formulation differs from that of \citep{Bick2016,Skardal2020HigherOI}
which derives from a systematic phase reduction near the Hopf bifurcation.

In this article, we report a generalized macroscopic property that
stems from the matrix coupling mechanism which we refer to as the
binary ``modes'' of the system. Given all possible values of the
coupling matrix in $\mathbb{R}^{d\times d}$, if the degree of coherence
is measured by each dimension, any dimension of the population has
the potential to synchronize through a second-order phase transition
when others remain incoherent, despite their explicit interdependence
in the model. Therefore, for a $d$ dimensional population, the system
is endowed with $2^{d}$ modes for every dimensional order parameter
to either take the value 0 or 1 in an approximate sense. The exact
combination of coherence or incoherence of the dimensions, along with
the derivation of the phase diagram, turns out to be closely linked
to the attributes of the coupling matrix. One of our main contributions
in this work is the necessary and/or sufficient conditions we established
through stability analysis, revealing how the algebraic properties
of the coupling matrix give rise to the system modes as a macroscopic,
statistical pattern. Additionally, we have observed explosive dimensional
phase transitions \citep{zhang2015explosive,Millan2019ExplosiveHK,Dai2020DiscontinuousTA,Skardal2020HigherOI}
in a numerical study demonstrated in the Supplementary Material, suggesting
that the phenomenon induced by this coupling mechanism may be far
richer than what is presented in the current study.

\section*{Results\label{sec:the model}}

\subsection*{Matrix coupling mechanism.}

The proposed dynamics reads

\begin{equation}
\dot{\theta}_{i}=\omega_{i}+\frac{1}{N}\cdot A\sum_{j=1}^{N}{\bf sin}(\theta_{j}-\theta_{i}),\label{eq:model}
\end{equation}

\noindent where $\theta_{i}=\left[\begin{array}{cccc}
\theta_{i1} & \theta_{i2} & ... & \theta_{id}\end{array}\right]^{T}\in\mathbb{R}^{d}$ denotes the multidimensional phase oscillator, each component of
$\theta_{i}$ is considered modulo $2\pi$ and has a natural frequency
specified by $\omega_{i}=\left[\begin{array}{cccc}
\omega_{i1} & \omega_{i2} & ... & \omega_{id}\end{array}\right]^{T}$. In this research, we focus on the heterogeneous population where
$\omega_{i}$ and $\omega_{j}$ can be distinct, and we assume they
follow a particular probability distribution. The function ${\bf sin}(\cdot):\mathbb{R}^{d}\rightarrow\mathbb{R}^{d}$
takes the sine of the operated vector dimension-wise, which is then
summed and weighed by a $d\times d$ real matrix $A$. To see an example,
a system of two-dimensional oscillators is expressed as
\begin{equation}
\dot{\theta}_{i}=\left[\begin{array}{c}
\dot{\theta}_{i1}\\
\dot{\theta}_{i2}
\end{array}\right]=\left[\begin{array}{c}
\omega_{i1}\\
\omega_{i2}
\end{array}\right]+\frac{1}{N}\cdot A\sum_{j=1}^{N}{\bf sin}\left(\left[\begin{array}{c}
\theta_{j1}-\theta_{i1}\\
\theta_{j2}-\theta_{i2}
\end{array}\right]\right)\label{eq:model-2d}
\end{equation}
where $\{\omega_{i1}\}$ and $\{\omega_{i2}\}$ are the natural frequencies
of oscillation on the $\{\theta_{i1}\}$ dimension and $\{\theta_{i2}\}$
dimension. Now, instead of the averaged scalar coupling in the standard
Kuramoto model, we consider an averaged, all-to-all coupling with
the matrix $A=\begin{bmatrix}a_{11} & a_{12}\\
a_{21} & a_{22}
\end{bmatrix}\in\mathbb{R}^{2\times2}$, that acts on the sum of vectors $[\sin(\theta_{j1}-\theta_{i1})\;\sin(\theta_{j2}-\theta_{i2})]^{T}$.
\begin{figure}
\begin{centering}
\includegraphics[width=12cm]{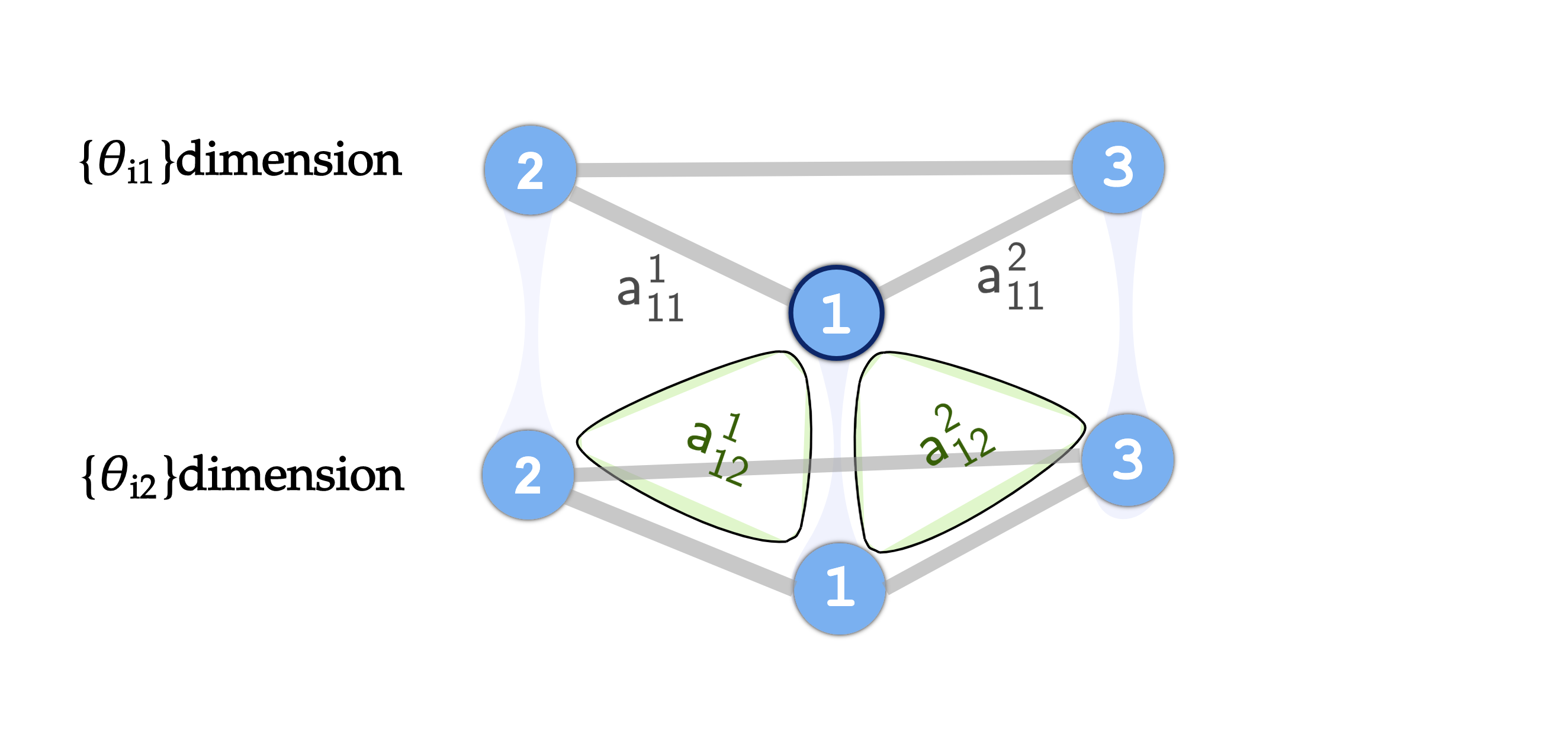}
\par\end{centering}
\caption{\emph{The matrix coupling mechanism}. Illustration of how the three
oscillators and their dimensions are interacting, with connection
between $\theta_{1}$ and $\theta_{2}$ weighed by $A_{1}=\begin{bmatrix}a_{11}^{1} & a_{12}^{1}\protect\\
a_{21}^{1} & a_{22}^{1}
\end{bmatrix}$; connection between $\theta_{1}$ and $\theta_{3}$ is weighed by
$A_{2}=\begin{bmatrix}a_{11}^{2} & a_{12}^{2}\protect\\
a_{21}^{2} & a_{22}^{2}
\end{bmatrix}$, and there is $A_{3}=\begin{bmatrix}a_{11}^{3} & a_{12}^{3}\protect\\
a_{21}^{3} & a_{22}^{3}
\end{bmatrix}$ for $\theta_{2}$ and $\theta_{3}$. Gray, solid lines indicate pairwise
interactions between components of the same dimension, which can also
be seen as the 1-simplex; the inter-dimensional interactions that
$\theta_{11}$ is involved in are marked with triangles, and can be
regarded as the 2-simplex. First-row elements of $A_{1},A_{2}$ weigh
the influences from these interactions on $\dot{\theta}_{11}$.}

\label{fig:FIG1}
\end{figure}
 FIG. \ref{fig:FIG1} demonstrates a simple connected network of $N=3$
where the interactions are weighed by matrices $A_{1},A_{2},A_{3}$,
with subscripts to avoid ambiguity. One should notice that the dynamics
on one dimension of a specific oscillator is taking direct influences
from both dimensions of the neighboring oscillators, except they are
weighed differently by the row elements of $A_{1}$ or $A_{2}$. If
we spell out the dynamics of $\theta_{11}$ in FIG. \ref{fig:FIG1},
there is 
\[
\dot{\theta}_{11}=\omega_{11}+\frac{1}{3}a_{11}^{1}\sin(\theta_{21}-\theta_{11})+\frac{1}{3}a_{12}^{1}\sin(\theta_{22}-\theta_{12})+\frac{1}{3}a_{11}^{2}\sin(\theta_{31}-\theta_{11})+\frac{1}{3}a_{12}^{2}\sin(\theta_{32}-\theta_{12}),
\]
where $A_{1}=(a_{ij}^{1}),A_{2}=(a_{ij}^{2})$. The expression contains
explicit pairwise terms that involve $\theta_{i1}$, namely $\frac{1}{3}a_{11}^{1}\sin(\theta_{21}-\theta_{11})+\frac{1}{3}a_{11}^{2}\sin(\theta_{31}-\theta_{11})$;
what make it difficult to connect this dynamics with real-world systems
are the rest, which do not concern $\theta_{i1}$. One way to resolve
this is to regard the terms from other dimensions as the result of
vector additions, e.g., $\frac{1}{3}a_{12}^{1}\sin(\theta_{22}-\theta_{12})=\frac{1}{3}a_{12}^{1}\sin\left(\left(\theta_{22}-\theta_{11}\right)-\left(\theta_{12}-\theta_{11}\right)\right)$
which implies a three-way interaction among $\theta_{11},\theta_{12}$,
and $\theta_{22}$. The matrix-coupled multidimensional variables
then display a simplicial complex structure that integrates intra-dimensional
links and inter-dimensional triangles. In essential, the dimensions
of the population become codependent under matrix multiplication,
and the intensity of this codependence is tuned by the $d^{2}$ matrix
elements.

The proposed dynamics would be further clarified if we define the
complex order parameter for each dimension as 
\begin{equation}
\rho_{r}=\frac{1}{N}\sum_{j=1}^{N}e^{i\theta_{jr}}=\sigma_{r}e^{i\Psi_{r}},r\in\{1,2,...,d\},\label{eq:def-orderpara}
\end{equation}
where the magnitude satisfies $0\leq\sigma_{r}\leq1$ and $\Psi_{r}$
denotes the average phase of $\theta_{jr}$ projected onto the complex
plane. When $\sigma_{r}\rightarrow1,$ it indicates a high degree
of synchrony on the $r$-th dimension of the oscillators, meanwhile
$\sigma_{r}\rightarrow0$ indicates an incoherent distribution of
$\{\theta_{jr}\}$ where their complex projections cancel out. In
our example of the two dimensional oscillators \eqref{eq:model-2d},
the equation of motion can then be adapted to

\begin{equation}
\begin{array}{c}
\dot{\theta}_{i1}=\omega_{i1}-a_{11}\sigma_{1}\sin(\theta_{i1}-\Psi_{1})-a_{12}\sigma_{2}\sin(\theta_{i2}-\Psi_{2}),\\
\dot{\theta}_{i2}=\omega_{i2}-a_{21}\sigma_{1}\sin(\theta_{i1}-\Psi_{1})-a_{22}\sigma_{2}\sin(\theta_{i2}-\Psi_{2}),
\end{array}\label{eq:4}
\end{equation}

\noindent which demonstrates that the instantaneous frequencies are,
in fact, directly controlled by the $\{\theta_{i1}\}$ mean-field
and $\{\theta_{i2}\}$ mean-field. The mean-field actions are then
modulated through the matrix elements of the same row.

In general, it is also possible to derive a compact form of \eqref{eq:model}
with the order parameters for $d$ dimensions, i.e., 
\begin{equation}
\dot{\theta}_{i}=\omega_{i}-A\Sigma{\bf sin}\left(\theta_{i}-\Psi\right).\label{eq:model-vec}
\end{equation}
Here $\Sigma={\bf diag}\left\{ \sigma_{1},\sigma_{2},...,\sigma_{d}\right\} ,\Psi=\begin{bmatrix}\Psi_{1} & \Psi_{2} & ... & \Psi_{d}\end{bmatrix}^{T}.$
Equation \eqref{eq:model-vec} is essential in determining the fixed
point of an individual and the distribution of the population. We
see from \eqref{eq:model-vec} that it is not always possible for
$\theta_{i}$ to be fixed on every dimension, where the natural frequency
$\omega_{ir}$ is dominating over the mean-field influence. When $\dot{\theta}_{i}=0$
is indeed solvable, it is ideal to have an invertible $A$ to obtain
a unique, closed-form expression of the fixed point $\theta_{i}^{*}$;
we refer to such oscillators as being fully synchronized (by the mean-field).
For oscillators that are not fixed on all of their dimensions, it
is expected that they will be relatively in motion to the entrained
population, for which we refer to them as the drifting oscillators.

Our main objective is to investigate the effect of varying coupling
matrices on the order parameter of each dimension, thus to identify
the qualitatively distinct states of the system induced by this coupling
mechanism. Now that the parameter space is augmented from the conventional
$\mathbb{R}$ to $\mathbb{R}^{d\times d}$, there are obviously numerous
ways that each element of the coupling matrix can be adjusted. Our
approach involves scaling an invertible, real symmetric initial matrix
by an incremental real number, commencing at zero. This enables us
to preserve, or at least trace, the algebraic characteristics of the
initial matrix while observing the complete transition the system
undergoes, from a state of no matrix coupling effect to one where
such coupling is particularly strong. We also make the assumption
that the natural frequency $\omega_{i}$ of an oscillator does not
distinguish between different dimensions, i.e., $\omega_{i}=\bar{\omega}_{i}{\bf 1}_{d}$,
whereas $\bar{\omega}_{i}$ for $i\in\{1,...,N\}$ follow a unimodal
probability distribution.

\subsection*{Binary system modes and independent dimensional phase transition.}

\begin{figure}
\begin{centering}
\includegraphics[viewport=0bp 0bp 927bp 239bp,width=18cm]{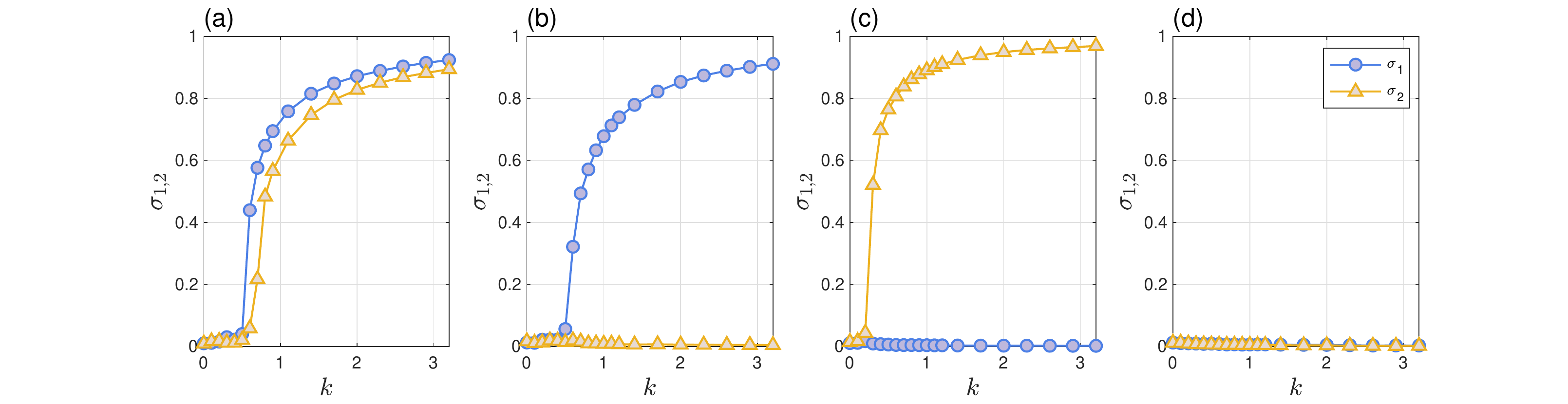}
\par\end{centering}
\caption{\emph{Transitions of $\sigma_{1,2}$ with initial matrices $A_{u}$.}
Panels (a), (b), (c), and (d) are results of direct simulation of
eqn. (\ref{eq:model-2d}) on $N=5000$ oscillators with $A=kA_{u},$
$A_{u}=A_{1},A_{2},A_{3},A_{4}$ respectively; each panel demonstrates
the values of the order parameters $\sigma_{1}$ (blue circle) and
$\sigma_{2}$ (yellow triangle) during the transition at each value
of $k$. Here $\sigma_{1}\rightarrow1(\sigma_{2}\rightarrow1)$ indicates
a high level of synchrony in dimension $\{\theta_{i1}\}(\{\theta_{i2}\})$
while $\sigma_{1}\rightarrow0(\sigma_{2}\rightarrow0)$ indicates
desynchrony, i.e., a uniform distribution in dimension $\{\theta_{i1}\}(\{\theta_{i2}\})$
which we observed from the numerical experiment.}

\label{fig:IDPT_2}
\end{figure}

\begin{figure}
\begin{centering}
\begin{tabular}[b]{llll}
\includegraphics[viewport=0bp 0bp 560bp 420bp,width=4cm]{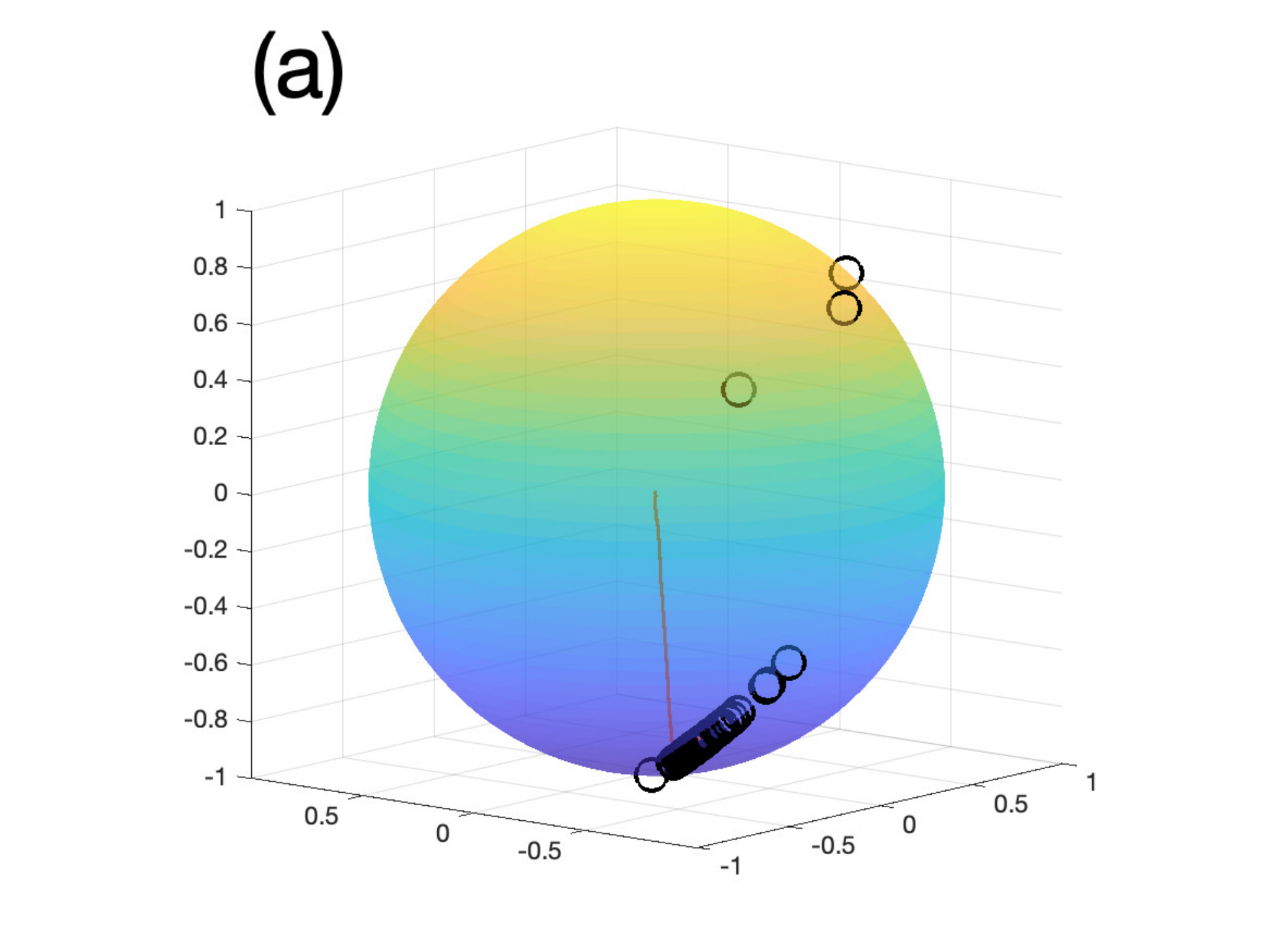} &
\includegraphics[viewport=0bp 0bp 560bp 420bp,width=4cm]{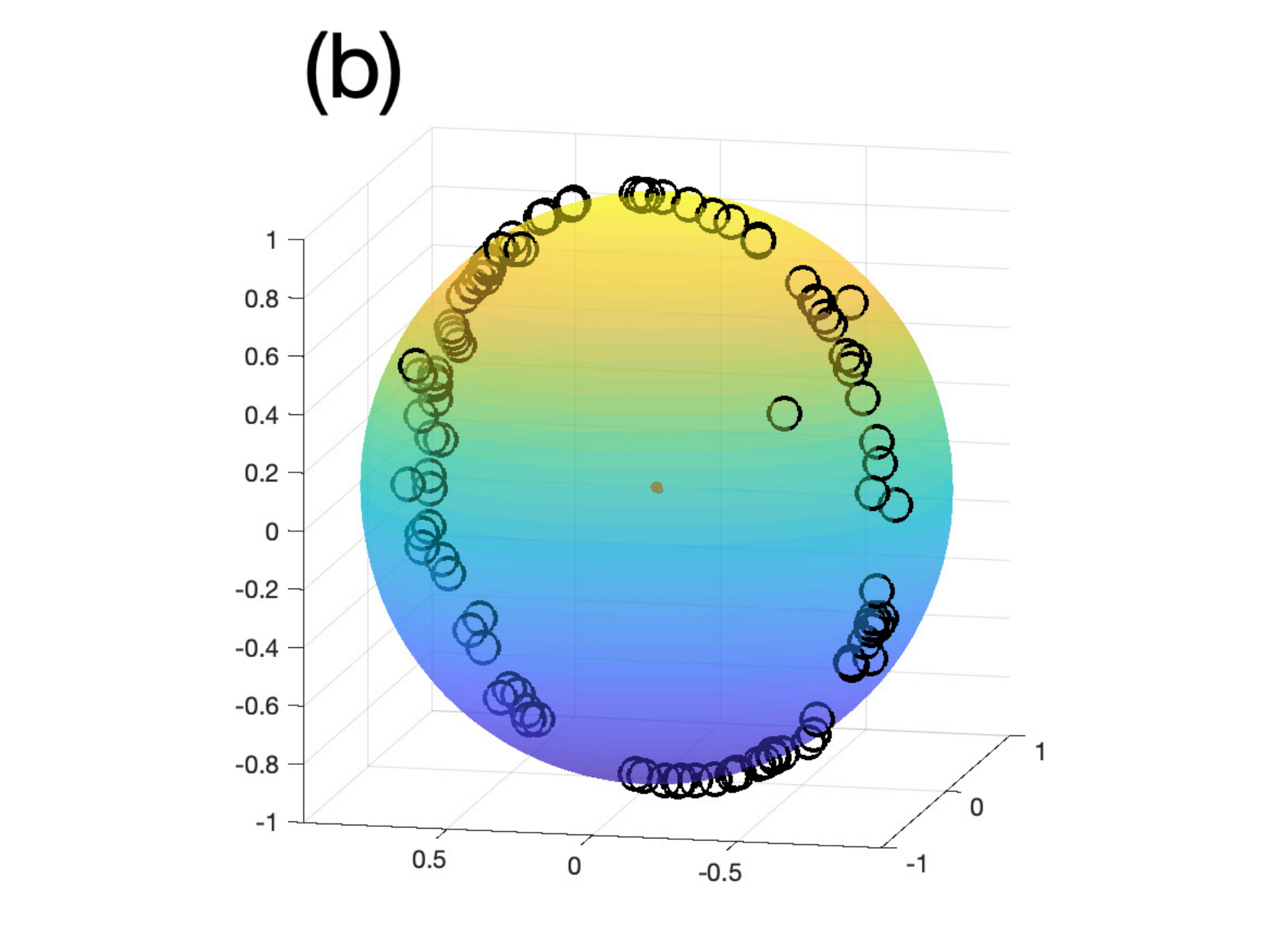} &
\includegraphics[viewport=0bp 0bp 560bp 420bp,width=4cm]{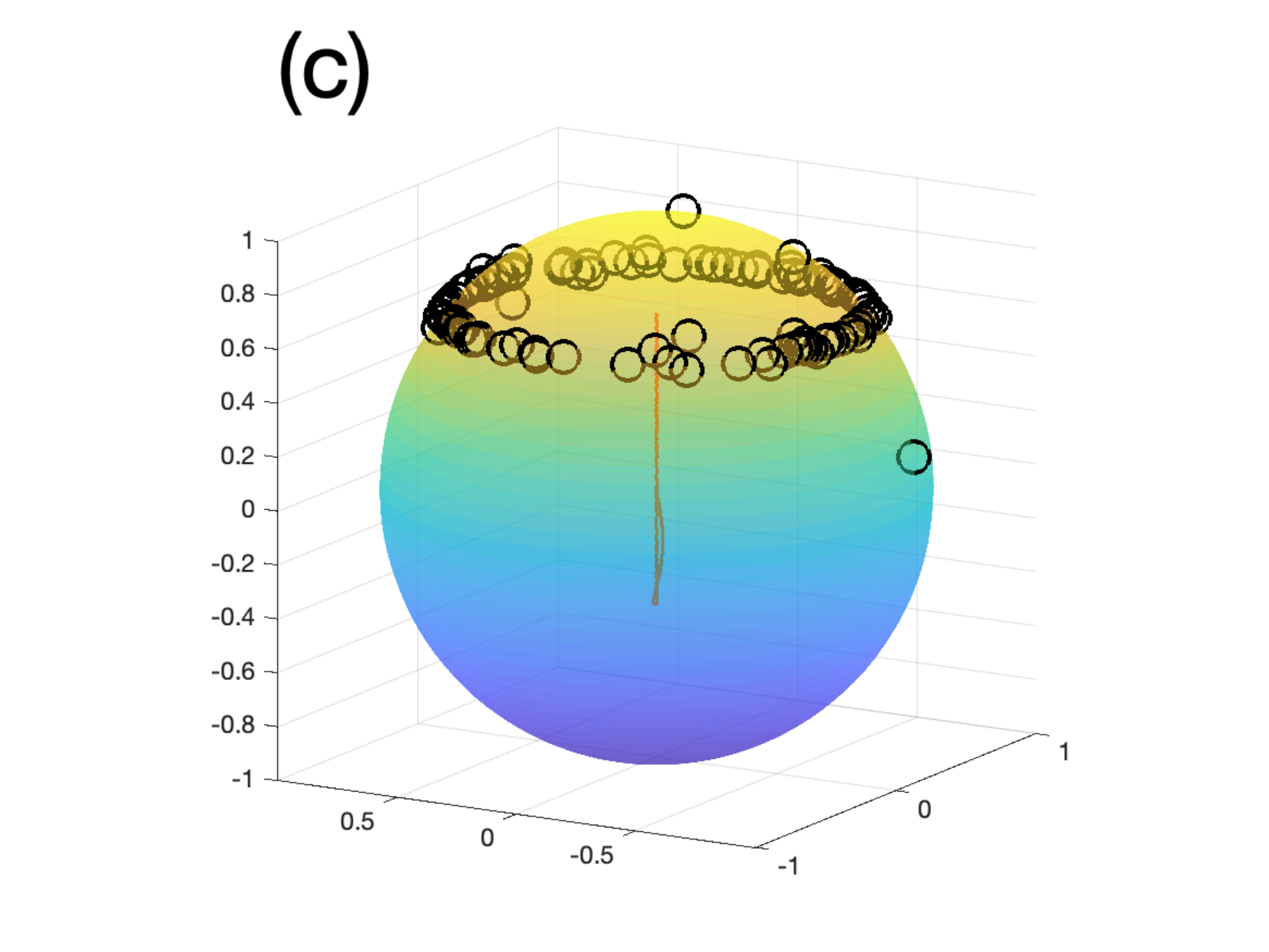} &
\includegraphics[viewport=0bp 0bp 560bp 420bp,width=4cm]{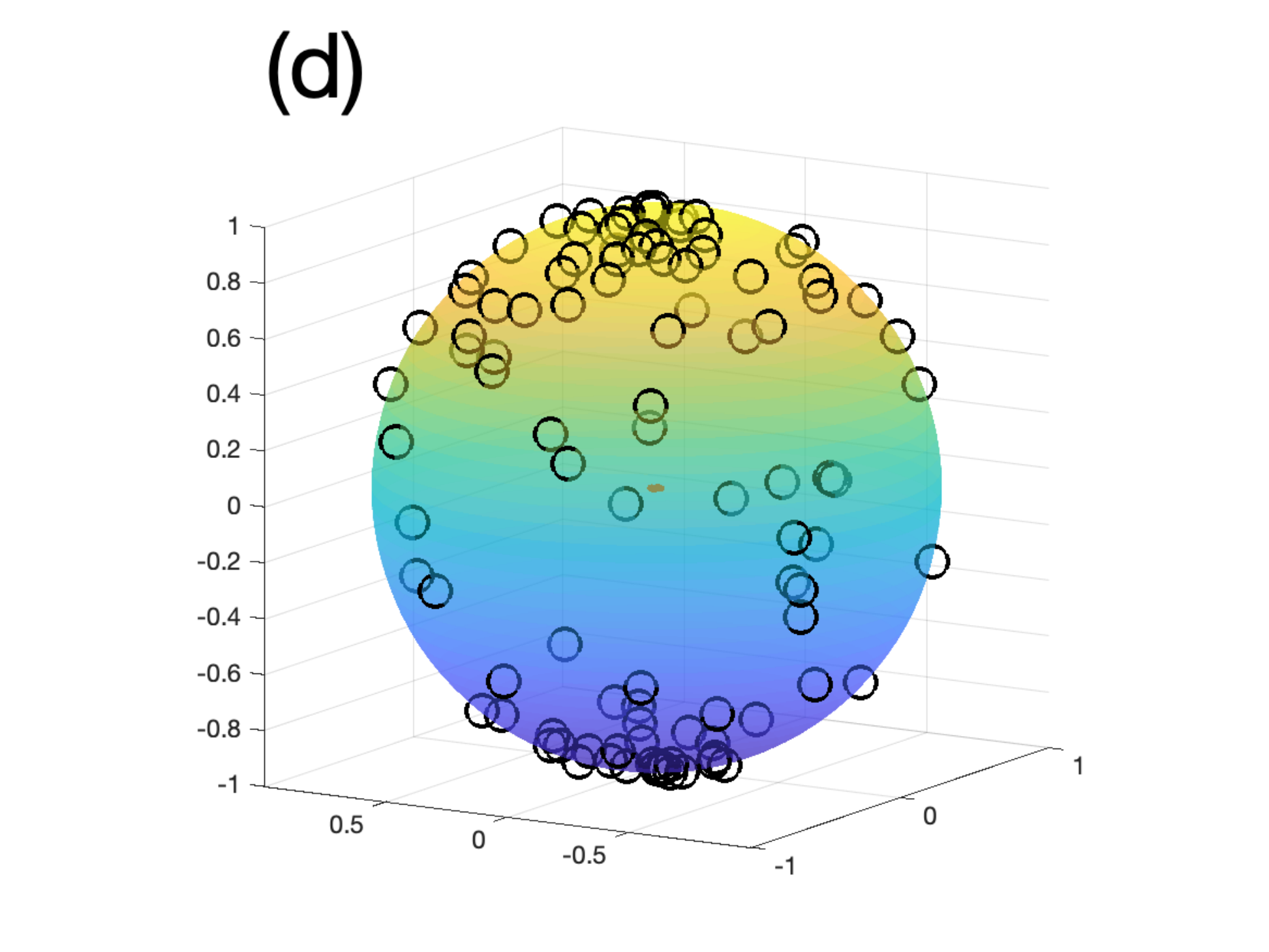}\tabularnewline
\end{tabular}
\par\end{centering}
\caption{\emph{Visualization of the system modes on the unit sphere.} Take
$\theta_{i1}$ as the azimuthal angle and $\theta_{i2}$ as the polar
angle, the population of oscillators (black circles) are projected
onto the surface of a unit sphere, where the orange line denotes the
center of the population for every time instant. We illustrate with
$100$ oscillators from the original ensemble $N=5000$ whose initial
conditions were uniformly distributed. The four panels correspond
to the four modes the system eventually settled in \textcolor{black}{with
the initial matrices $A_{1},A_{2},A_{3},A_{4}$} at $k=10$. Panel
(a) displays mode 11 where almost all oscillators are fully synchronized
(in frequency) for both elements $\theta_{i1}$ and $\theta_{i2}$;
the oscillators remain relatively static and travel at a constant,
but minuscule speed of the mean field. Panel (b) shows, at mode 10,
the oscillators are only synchronized on the $\theta_{i1}$ element
and stay disarranged on the ring that rotates slowly about the $z$-axis,
their average position soon approaching the origin. Panel (c) corresponds
to mode 01 where the population is only synchronized on the $\theta_{i2}$
element and is distributed on a ring that is shrinking and stretching
as it moves up and down in the $z$-direction. Panel (d) demonstrates
how the population is incoherent on both dimensions at mode 00, where
oscillators drift across the surface and do not form any groups or
show any particular pattern. }

\label{fig:IDPT_sphere}
\end{figure}

Under the above assumptions, the main finding of this paper is demonstrated
in FIG. \ref{fig:IDPT_2} from an experiment we performed on $N=5000$
oscillators. The heterogeneous population is mediated respectively
by the four randomly generated matrices $A_{1},A_{2},A_{3},A_{4}$
from a uniform initial distribution (see Method for details). In each
case of $A=kA_{u},u\in\{1,2,3,4\}$, we adiabatically increase $k$
from $0$ to $3.2$ in 22 steps and calculate the average of $\sigma_{1,2}$
in a span of time, when the order parameters are considered to have
reached their equilibrium. We note that this scaling is fundamentally
different from the strengthening $K$ in other generalized Kuramoto
models, as the matrix elements are inherently $d(d+1)/2$ independent
variables undergoing changes in the parameter space. As a result,
FIG. \ref{fig:IDPT_2} reports that with $A_{1}$ and $A_{4}$, both
the $\{\theta_{i1}\}$ dimension and the $\{\theta_{i2}\}$ dimension
go through qualitatively similar transitions either from being largely
incoherent, due to the limit size effect, to a complete frequency
synchronization ($\sigma_{1,2}\rightarrow1$), or to a complete desynchronization
($\sigma_{1,2}\rightarrow0$). Meanwhile with $A_{2}$ and $A_{3}$,
the transitions on $\{\theta_{i1}\}$ dimension and $\{\theta_{i2}\}$
dimension go to opposite directions, i.e., $\sigma_{1}\rightarrow1,\sigma_{2}\rightarrow0$
or $\sigma_{1}\rightarrow0,\sigma_{2}\rightarrow1$ as the elements
of $A$ are simultaneously and sufficiently increased. Take $\theta_{i1}$
as the azimuthal angle and $\theta_{i2}$ as the polar angle; though
there is a slight abuse of this coordinate system as $\theta_{i2}$
deviates from the conventional $[0,\pi]$, we can project these combinations
of $\sigma_{1,2}$ onto the unit sphere for visualization, i.e., $x_{i}=\left[\begin{array}{ccc}
\sin\theta_{i2}\cos\theta_{i1} & \sin\theta_{i2}\sin\theta_{i1} & \cos\theta_{i2}\end{array}\right]^{T}$. FIG. \ref{fig:IDPT_sphere} demonstrates how the system has set
into four qualitatively distinct modes of distribution and motion,
each to a configuration of $A_{u}$. In fact, FIG. \ref{fig:IDPT_sphere}
validates our definition of a set of dimensional order parameters
$\{\rho_{1},\rho_{2},...,\rho_{d}\}$, because in the thermodynamic
limit $N\rightarrow\infty$, the mean field on the synchronized dimension
is supposed to lose its velocity due to the symmetrically distributed
$\bar{\omega}_{i}$, and thus, using the definition $\left|\rho\right|=\left|\frac{1}{N}\sum_{i=1}^{N}x_{i}\right|$,
we see that the $\sigma_{1}\sigma_{2}=01$ mode will be indicated
by a $\left|\rho\right|$ that settles into a constant between 0 and
1, while mode $\sigma_{1}\sigma_{2}=10$ and mode $\sigma_{1}\sigma_{2}=00$
are indistinguishable given $\left|\rho\right|=0$. None of the above
values of $\left|\rho\right|$ asserts that certain parts of the multidimensional
$\theta_{i}$ are in fact synchronized, i.e., there is lost information
in measuring the system with a sole indicator, when we should be benefiting
from a generalization that entails individuals inherently carrying
more information. 

\begin{figure}
\begin{centering}
\includegraphics[viewport=0bp 0bp 774bp 366bp,width=10cm]{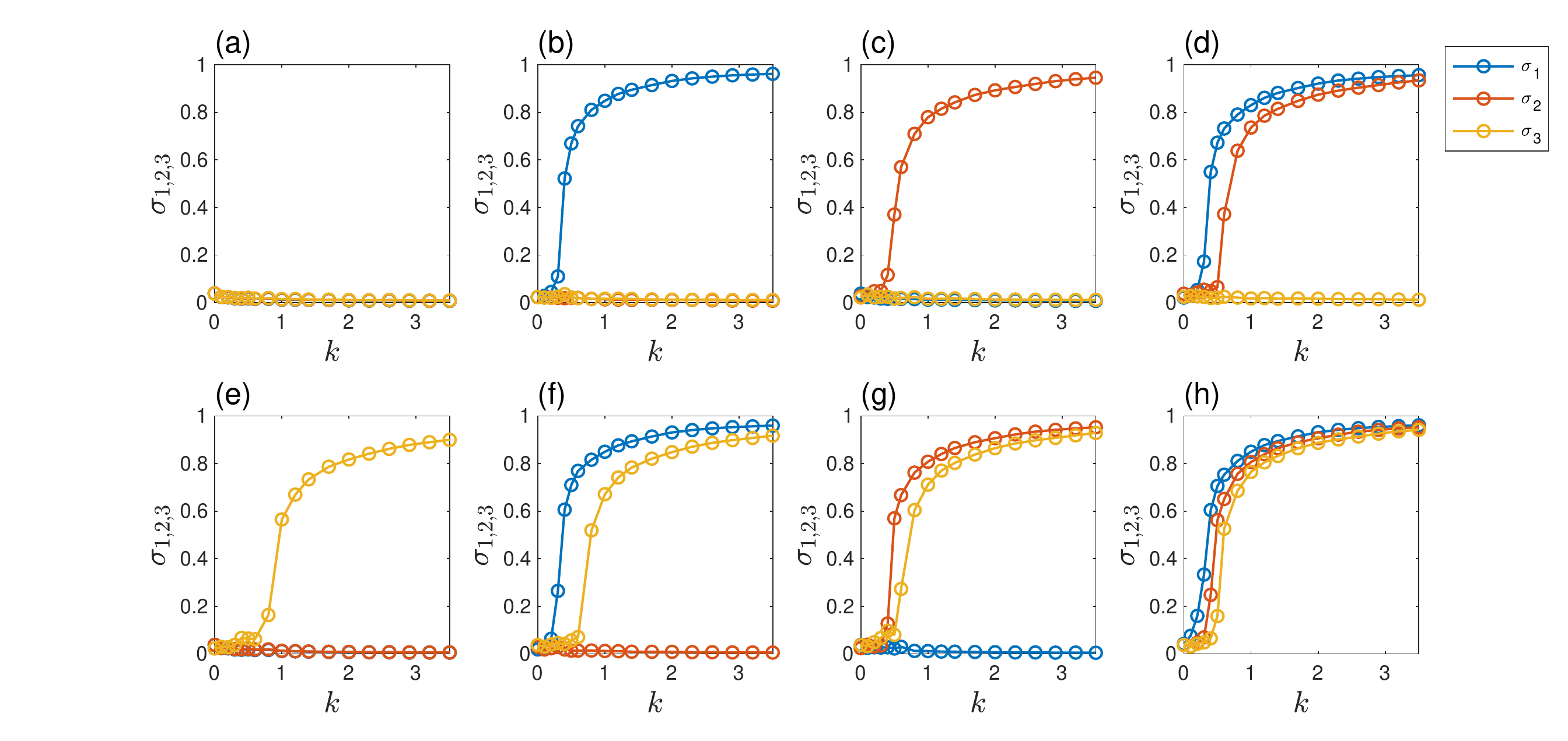}\includegraphics[width=9cm]{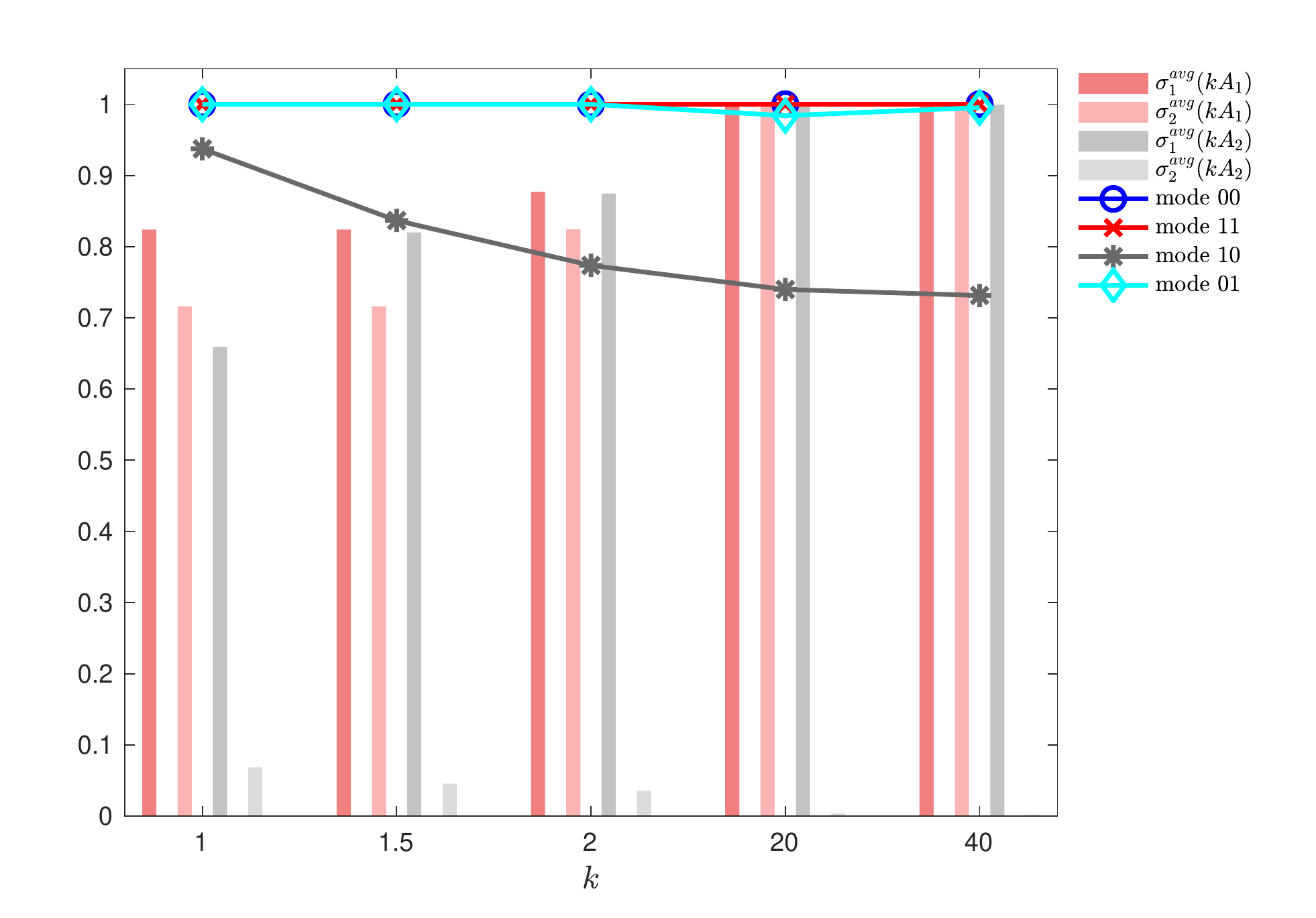}
\par\end{centering}
\caption{\emph{Transition of $\sigma_{1,2,3}$ for the three-dimensional oscillators
and the initial value experiment}. Left: panels (a)$\sim$(h) are
simulation results of eqn. (\ref{eq:model-2d}) on $N=1000$ three-dimensional
oscillators with coupling matrices $A_{i}=U^{T}diag\{5s_{1}^{i},6s_{2}^{i},8s_{3}^{i}\}U,i=1,...,8$
where $U=\left[\protect\begin{array}{ccc}
-0.2619 & -0.3259 & -0.9084\protect\\
0.0652 & -0.9451 & 0.3202\protect\\
-0.9629 & 0.0246 & 0.2688
\protect\end{array}\right]$ and $(s_{1}^{i},s_{2}^{i},s_{3}^{i})$ runs from $(-1,-1,-1)$ to
$(+1,+1,+1)$ in a combinatorial sense, the corresponding system modes
$\sigma_{3}\sigma_{2}\sigma_{1}$ also runs from $000$ (panel (a))
to $111$ (panel (h)). Right: line graph illustrates the percentage
of the $10^{4}$ uniformly distributed initial values that converge
to the same system mode at $kA_{u}$ to that in FIG. \ref{fig:IDPT_2},
for $u\in\{1,2,3,4\},k\in\{1,1.5,2,20,40\}$. Bar chart demonstrates
the average values of $\sigma_{1},\sigma_{2}$ at each $k$ that are
considered to have chosen mode 11 with $A_{1}$ (red bars), and those
considered to have chosen mode 10 with $A_{2}$ (gray bars). The complete
data are demonstrated in the Supplementary Material.}

\label{fig:IDPT_3}
\end{figure}

Motivated by what is observed from the two-dimensional population,
we then apply the same method on an ensemble of three dimensional
Kuramoto oscillators, coupled through matrices $A_{u}\in\mathbb{R}^{3\times3}$,
producing a total of eight combinations of $\sigma_{1,2,3}$ as depicted
in FIG. \ref{fig:IDPT_3}. 
 At this stage, what we have shown with two dimensional and three
dimensional population is that, the matrix coupling enables the multidimensional
Kuramoto oscillators to separate the transition to coherence/incoherence
on each dimension, which we call an independent dimensional phase
transition.

However, among the majority of the generalized Kuramoto models, it
is not uncommon to have systems that are multistable at a particular
coupling strength or within a range of the coupling strength; therefore
the basins of attraction become decisive in the emergent equilibria
\citep{Wiley2006TheSO,Delabays2017TheSO}. In FIG. \ref{fig:IDPT_3},
our experiment on $N=100$ matrix-coupled oscillators indicates that,
with a fixed value of the matrix $A=kA_{u}$, the system modes may
indeed be limitedly attracting to the $10^{4}$ uniformly distributed
initial values depending on $k$. For the criteria that determine
which system mode the data correspond to, even when $\sigma_{r}$
may be significantly between $(0,1)$ due to small $k$, we refer
to the content in Method.

Notably, in our experiment with $A_{2}$, a lower value of $k$ in
general results in a higher percentage of consistency with the phase
transition results in FIG. \ref{fig:IDPT_sphere}. Yet interestingly,
when we gather the 2668 samples of $\theta(0)$ that converge to mode
01 at $k=20$, and start to progressively increase $k$ from zero
with these initial values, the dimensional phase transitions induced
are all towards the equilibrium mode 10 as those in FIG.(3b). In fact,
all $10^{4}$ initial values have yielded the above result with the
adiabatically increased $k\in[0,2.3]$. The same occurred to our examination
of the 160 samples that converge to mode 10 with $kA_{3},k=20$, where
the system eventually converges to mode 01 with incremental $k$,
as all $10^{4}$ initial values do. This suggests that a weak matrix-coupling
favors a particular combination of dimensional phase transitions at
an incoherent state of the system, so that it is dominantly attracting
for the vast possibilities of initial values; while this might not
be the case when the matrix effect is particularly strong. We also
confirms that the phenomenon appeared in FIG. \ref{fig:IDPT_sphere}
is, to some extent, robust to adjustments in initial values, and the
choice of the coupling matrix is indeed relevant to the corresponding
system mode that emerges at the end of the transition.

\subsection*{Summary of the theoretical results.}

Our theoretical analysis addresses several key aspects of the $d$-dimensional
population: the steady state solutions to the order parameters, analytical
estimate of the onset of dimensional phase transition, and the relation
between the coupling matrix and the system mode. Under the premise
that $A$ is real symmetric and invertible, we have derived the following
conclusions on the last question based on stability analysis:

(1) The system mode that corresponds to a full coherence ($\sigma_{1}\sigma_{2}\cdots\sigma_{d}\rightarrow11\cdots1$)
is linearly stable at the end of the transition $A=kA_{initial}$
if and only if the initial matrix is positive definite.

(2) If the system admits mode $\sigma_{i_{1}}\sigma_{i_{2}}\cdots\sigma_{i_{\tau}}\sigma_{i_{\tau+1}}\sigma_{i_{\tau+2}}\cdots\sigma_{i_{d}}\rightarrow\underset{\tau}{\underbrace{\begin{array}{cccc}
0 & 0 & \cdots & 0\end{array}}}\underset{d-\tau}{\underbrace{\begin{array}{cccc}
1 & 1 & \cdots & 1\end{array}}}$ at the end of the transition $A=kA_{initial}$, then the primary
submatrix consisting dimensions $\{i_{\tau+1},i_{\tau+2},...,i_{d}\}$
of $A$ is positive definite.

(3) If there exists a diagonal element $a_{rr}<0$, then $\sigma_{r}\rightarrow0$,
the $r$-th dimension of the population will remain incoherent for
the entire transition.

(4) If the initial matrix is negative definite, then the incoherent
solution is stable on every dimension, and we observe $\sigma_{1}\sigma_{2}\cdots\sigma_{d}\rightarrow00\cdots0$.

(5) Given a $d$ dimensional population, the $2^{d}$ system modes
exist under the matrix coupling mechanism.

\noindent Here (1) and (2) are obtained through a linearization of
the system of finite population around the fully coherent solution,
followed by our analysis on the distribution of the Jacobian eigenvalues;
(4) is an inference from (3) given that $A$ is real symmetric, both
of which stem from our analysis on the linear stability of the fully
incoherent solution in the thermodynamic limit. To prove statement
(5), consider statements (1) and (3) and an arbitrary combination
of $\sigma_{1}\sigma_{2}\cdots\sigma_{d}$, where $\bar{\mathbb{S}}=\{i_{1},i_{2},...,i_{\tau}\}\subset\{1,2,...,d\}$
denotes dimensions that are incoherent, and $\mathbb{S}=\{i_{\tau+1},i_{\tau+2},...,i_{d}\}$,
coherent. If we construct $A_{initial}$ in such a way that $a_{rr}<0$
for $r\in\bar{\mathbb{S}}$, while the reduced system that leaves
out these dimensions, expressed by equation \eqref{eq:model_reduced},
is characterized by an $A'$ that is rid of the corresponding rows
and columns to $r\in\bar{\mathbb{S}}$. Then by ensuring $A'$ to
be positive definite, we know for $r\in\mathbb{S}$ there is coherence
due to (1).

Since the calculation of the coherent branch of $\sigma_{r}$ does
not provide a clear, analytical expression of the bifurcation point
$A_{critical}=k_{critical}A_{u}$, we have adopted the method in Ref.
\citep{strogatz2000kuramoto} and set up a multi-variable Fourier
formulation, which yields similar characteristic equations on the
eigenmodes to that of the classic model, except the coupling strength
that determines the stability of incoherence is now replaced by the
diagonal elements of the matrix (see Method). The characteristic equations
then predict that, should there be a phase transition towards synchronization
on $\sigma_{r}$, it would happen when the diagonal element $a_{rr}$
is scaled across a critical value of $\frac{2}{\pi g(0)}$. Considering
the diagonal elements are not necessarily identical, the bifurcation
of each $\sigma_{r}$ understandably occurs at a distinct stage of
the scaling, hence the differences in $k_{critical}$ for $\sigma_{1,2}$
in FIG. \ref{fig:IDPT_sphere}, and for $\sigma_{1,2,3,}$ in FIG.
\ref{fig:IDPT_3}. Note that the dimension with the larger diagonal
is earlier in reaching the bifurcation point. 

\begin{figure}
\begin{centering}
\includegraphics[width=12cm]{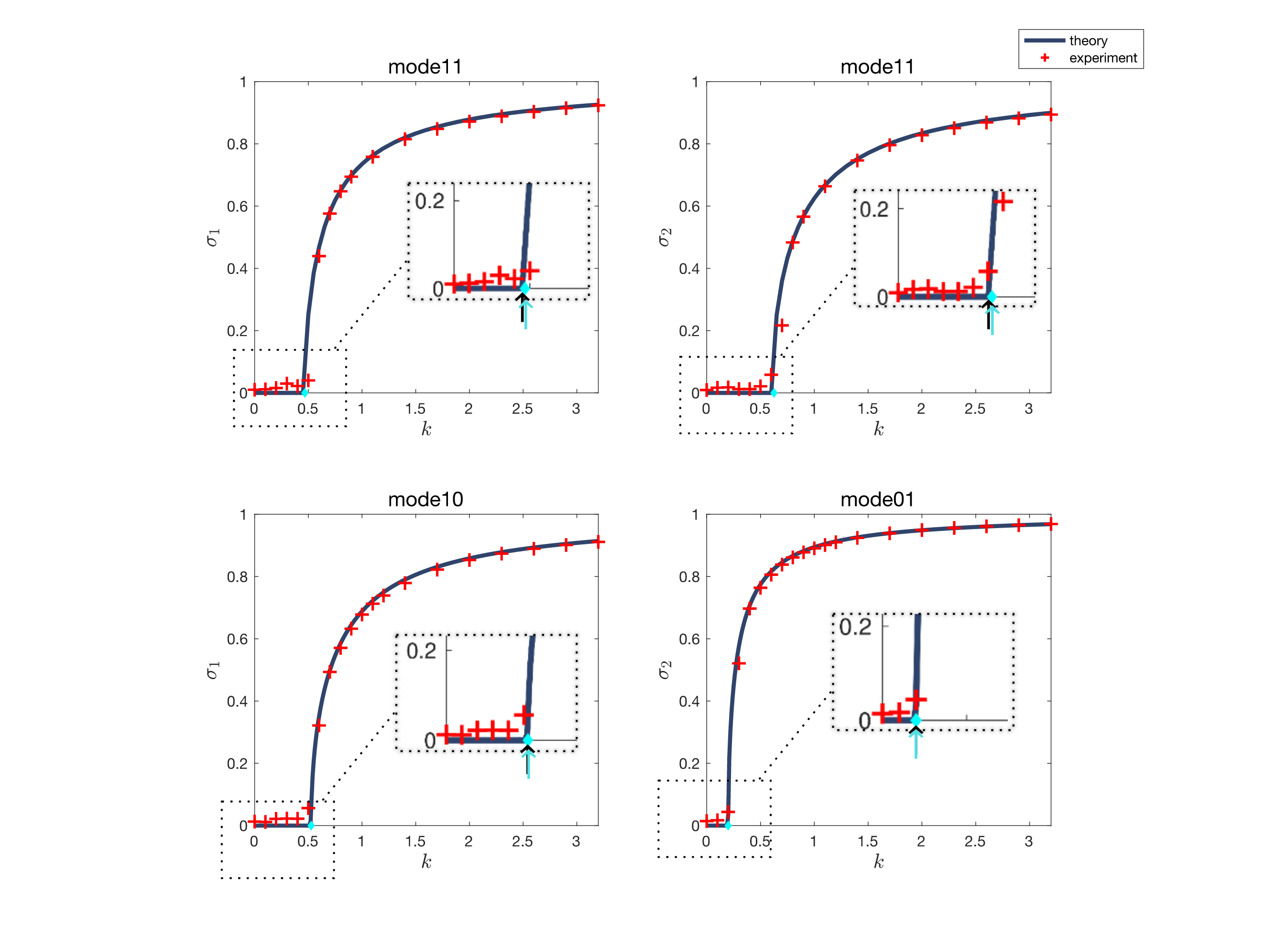}
\par\end{centering}
\caption{\emph{The steady state solutions and simulation data of dimensional
phase transition}s. The four panels demonstrate the order parameter
magnitudes $\sigma_{1,2}$ undergoing phase transitions under mode
11, mode 10, and mode 01. In the insets, we compared the critical
value of $k$ from the data and that predicted by the self-consistent
analysis and the stability analysis (marked with cyan diamond). The
steady state solutions are obtained by solving equations \eqref{eq:sigma_classic}
and \eqref{eq:sigma_12} using Matlab, fsolve. For eqn. \eqref{eq:sigma_12},
the initial evaluation of the equations is set as $\sigma_{1}=\sigma_{2}=0.3$
for $k\leqslant0.55$, while it approaches the experiment data for
$k>0.55$.}

\label{fig:FIG5}
\end{figure}

A comparison of the simulation result and the theoretical prediction
of the dimensional phase transition is presented in FIG. \ref{fig:FIG5}.
Notice the close fit between the analytically derived coherent $\sigma_{r}$
and the simulation data on the $N=5000$ population. We have also
marked up predictions of $k_{critical}$ from the steady state solution
and the stability analysis in the insets. The theoretical predictions
show the best agreement with the data when only one of the two dimensions
is synchronizing, reducing the system to the classic Kuramoto model.
When more than one dimensions is undergoing phase transition, both
predictions of $k_{critical}$ slightly deviate from the data and
also from each other. 

\subsection*{Linear Stability of the Coherent Solution.}

Let us first present the theoretical ground on statement (1). Before,
we made the observation that when the weight matrix $A$ is real symmetric,
a full synchronization emerges at a positively distributed spectrum
of $A$. This has been observed in oscillators of dimension two as
well as dimension three. By linearizing the whole system around the
coherent solution ($\sigma_{r}\rightarrow1,r=1,...,d$), we will show
that this correspondence of $\lambda(A)$ to the order parameters
is readily generalized into arbitrary dimension $d$. To see this,
consider the $Nd$ dimensional dynamic system \eqref{eq:model}. To
perform the linear stability analysis for the entire system, we would
like to evaluate the Jacobian matrix with respect to the state variable
$\boldsymbol{\theta}$ in a closed form. This is realized by identifying

\begin{align*}
\frac{\partial\dot{\theta}_{ir}}{\partial\theta_{ir}} & =-\frac{1}{N}a_{rr}\sum_{j}\cos(\theta_{jr}-\theta_{ir}), & \frac{\partial\dot{\theta}_{ir}}{\partial\theta_{jr}} & =\frac{1}{N}a_{rr}\cos(\theta_{jr}-\theta_{ir}),\\
\frac{\partial\dot{\theta}_{ir}}{\partial\theta_{is}} & =-\frac{1}{N}a_{rs}\sum_{j}\cos(\theta_{js}-\theta_{is}), & \frac{\partial\dot{\theta}_{ir}}{\partial\theta_{js}} & =\frac{1}{N}a_{rs}\cos(\theta_{js}-\theta_{is}),
\end{align*}
with $i\neq j\in\{1,...,N\},r\neq s\in\{1,...,d\}$, and $a_{rr},a_{rs}$
are the elements of the coupling matrix $A$. Then by sequencing $\boldsymbol{\theta}$
dimension-wise as $[\theta_{11},\theta_{21},...,\theta_{N1},...,\theta_{1d},\theta_{2d},...,\theta_{Nd}]$,
the Jacobian matrix can be written as 
\begin{equation}
J(\boldsymbol{\theta})=-\frac{1}{N}\begin{bmatrix}a_{11}\mathcal{L}_{1} & a_{12}\mathcal{L}_{2} & \cdots & a_{1d}\mathcal{L}_{d}\\
a_{21}\mathcal{L}_{1} & a_{22}\mathcal{L}_{2} & \cdots & a_{2d}\mathcal{L}_{d}\\
\vdots & \vdots & \ddots & \vdots\\
a_{d1}\mathcal{L}_{1} & a_{d2}\mathcal{L}_{2} & \cdots & a_{dd}\mathcal{L}_{d}
\end{bmatrix}=-\frac{1}{N}\left(A\otimes I_{N}\right)\begin{bmatrix}\mathcal{L}_{1} &  &  & O\\
 & \mathcal{L}_{2}\\
 &  & \ddots\\
O &  &  & \mathcal{L}_{d}
\end{bmatrix},\label{eq:J}
\end{equation}
where $\mathcal{L}_{r},r\in\{1,...,d\}$ are the graph Laplacians
of networks $\mathcal{G}_{r}$ for the $r$-th dimension of the population.
Because of \eqref{eq:model}, the networks $\mathcal{G}_{r}=(\mathcal{V}_{r},\mathcal{E}_{r},\mathcal{W}_{r})$
are (i) complete, (ii) undirected, and (iii) weighed by $\cos(\theta_{jr}-\theta_{ir})$
for any edge $\{ir,jr\}\in\mathcal{E}_{r}$. Formalizing this with
the incidence matrix $H$ for the complete graph, there is
\begin{align*}
\mathcal{L}_{r} & (\boldsymbol{\theta})=H{\bf diag}\{\cos(\theta_{jr}-\theta_{ir})\}H^{T}.
\end{align*}
When $\left|\theta_{jr}-\theta_{ir}\right|<\frac{\pi}{2}$, $\mathcal{G}_{r},r\in\{1,...,d\},$
are weighed by strictly positive scalars that render the Laplacians
$\mathcal{L}_{r}$ positive semi-definite, with $\text{nullity}(\mathcal{L}_{r})=1$. 

Let $P=A\otimes I_{N},Q={\bf blkdiag}\{\mathcal{L}_{1},\mathcal{L}_{2},...,\mathcal{L}_{d}\}$
and $J(\boldsymbol{\theta})=-\frac{1}{N}PQ$, we denote the eigenvalues
of $A\in\mathbb{R}^{d\times d}$ as $\lambda_{1},\lambda_{2},...,\lambda_{d}$,
and that of $Q$ as $\upsilon_{1},\upsilon_{2},...,\upsilon_{Nd}$,
while the Jacobian has eigenvalues $\mu_{1},\mu_{2},...,\mu_{Nd}$.
We will first look into some of the properties of the eigenvalues
of $P,Q$ respectively, before studying their combined effect on the
product of $P,Q$, which leads to the distribution of $\mu_{1},\mu_{2},...,\mu_{Nd}$
and an explanation of the independent dimensional transition to synchronization.

We first notice that for $P=A\otimes I_{N}$, its eigenvalues are
$\underset{N}{\underbrace{\begin{array}{cccc}
\lambda_{1}, & \lambda_{1} & ,\ldots, & \lambda_{1}\end{array}}},\ldots,\underset{N}{\underbrace{\begin{array}{cccc}
\lambda_{d}, & \lambda_{d} & ,\ldots, & \lambda_{d}\end{array}}}.$ Since $A$ is assumed to be Hermitian and invertible, $P$ is also
Hermitian, with similarly distributed positive or negative eigenvalues
of multiplicity $N$. The block diagonal matrix $Q$, on the other
hand, is clearly Hermitian, and its eigenvalues $\mu_{1},\mu_{2},...,\mu_{Nd}$
are that of the graph Laplacians. But for now we have not established
that $\mathcal{L}_{1},\mathcal{L}_{2},...,\mathcal{L}_{d}$ are positive
semidefinite, because $\left|\theta_{jr}-\theta_{ir}\right|<\frac{\pi}{2}$
may not hold in general. Consider the model \eqref{eq:model-vec},
if $\theta_{i}$ has a fixed point $\theta_{i}^{*}$, then 
\[
{\bf sin}\left(\theta_{i}^{*}-\Psi\right)=\Sigma^{-1}A^{-1}\omega_{i},
\]
it means that the condition that $\theta_{i}^{*}$ exists is 
\begin{equation}
\left\Vert \Sigma^{-1}A^{-1}\omega_{i}\right\Vert _{\infty}\leqslant1.\label{eq:fixpt-ineq}
\end{equation}
Note that $\Sigma^{-1}$ exists because we are going to evaluate the
stability of the coherent solution. In our experiment, the natural
frequency vector $\omega_{i}$ is set to be $\bar{\omega}_{i}{\bf 1}_{d}$,
and the coupling matrix $A$ is scaled from an initial matrix $A_{u}$
by a factor $k$. Therefore \eqref{eq:fixpt-ineq} is also
\begin{equation}
\left|\frac{\bar{\omega}_{i}}{k}\right|\left\Vert \Sigma^{-1}A_{u}^{-1}{\bf 1}_{d}\right\Vert _{\infty}\leqslant1.
\end{equation}
One sees that with finite $N$, as long as $k$ is large enough, the
fixed point $\theta_{i}^{*}$ always exists. Moreover, for oscillators
$\theta_{i}$ and $\theta_{j}$, 
\begin{align*}
\theta_{ir}^{*} & =\Psi_{r}+\sin^{-1}\left(\left(\Sigma^{-1}A^{-1}\omega_{i}\right)_{r}\right),\\
\theta_{jr}^{*} & =\Psi_{r}+\sin^{-1}\left(\left(\Sigma^{-1}A^{-1}\omega_{j}\right)_{r}\right),
\end{align*}
thus
\[
\theta_{ir}^{*}-\theta_{jr}^{*}=\sin^{-1}\left(\left|\frac{\bar{\omega}_{i}}{k}\right|\left(\Sigma^{-1}A_{u}^{-1}{\bf 1}_{d}\right)_{r}\right)-\sin^{-1}\left(\left|\frac{\bar{\omega}_{j}}{k}\right|\left(\Sigma^{-1}A_{u}^{-1}{\bf 1}_{d}\right)_{r}\right).
\]
Again, for large enough $k$ there will be $\left|\theta_{ir}^{*}-\theta_{jr}^{*}\right|<\frac{\pi}{2}$
for arbitrary $i,j\in\{1,...,N\},i\neq j$, and for arbitrary dimension
$r\in\{1,...,d\}$. Therefore, the graph Laplacians $\mathcal{L}_{r},r\in\{1,...,d\}$
are positive semi-definite, so is the block diagonal matrix $Q$.
We now have a simple bound for $\upsilon_{1},\upsilon_{2},...,\upsilon_{Nd}$
that is $\upsilon\geqslant0$.

For the Jacobian matrix $J(\boldsymbol{\theta})=-\frac{1}{N}PQ$ whose
eigenvalues are denoted as $\mu_{1},\mu_{2},...,\mu_{Nd}$, we mention
that these eigenvalues are real when $P,Q$ satisfy the above conditions.
This is easy to prove considering $PQ$ and $QP$ have the same non-zero
eigenvalues. Suppose $u$ is an eigenvector of $QP$ with respect
to eigenvalue $\mu_{i}'=-N\mu_{i}$, then we have $\left\langle QPu,Pu\right\rangle =\mu_{i}'\left\langle u,Pu\right\rangle \geqslant0$
because $Q\succeq0$; this also suggests that $\mu_{i}'\left\langle u,Pu\right\rangle $
is real. Then for $\mu_{i}'\neq0$, since $P$ is Hermitian, $\left\langle u,Pu\right\rangle $
is real, we have that $\mu_{i}',i=1,...,dN-d$ are real, so are $\mu_{1},\mu_{2},...,\mu_{Nd}$.
To further specify the distribution of the Jacobian eigenvalues, it
is necessary to introduce the following lemma from \citep[Theorem 1]{Ostrowski1959berEV}.
\begin{lem}
\label{lem:1}Let $A$ be a positive definite or semidefinite Hermitian
matrix of order $n$ with the smallest eigenvalue $m$ and the largest
eigenvalue $M$. Let $B$ be a Hermitian matrix of order $n$ with
eigenvalues $\Pi_{1}\leq\Pi_{2}\leq...\leq\Pi_{n}$. Then $AB$ has
real eigenvalues $\Lambda_{1}\leq...\leq\Lambda_{n}$, and it holds
for suitable factors $\Theta_{v}$, which lie between $m$ and $M$,
that
\begin{equation}
\Lambda_{v}=\Theta_{v}\Pi_{v},m\leq\Theta_{v}\leq M,(v=1,...,n).
\end{equation}
\end{lem}
Since both $P$ and $Q$ are square, $PQ$ actually has exactly the
same eigenvalues as $QP$, for which $Q$ is positive semi-definite,
$P$ is Hermitian and invertible. Let the eigenvalues of $PQ$ and
$QP$ be sequenced as $\mu_{1}'\leqslant\mu_{2}'\leqslant...\leqslant\mu_{Nd}'$.
We know that $\text{rank}(QP)=\text{rank}(Q)=Nd-d,$ which means there
are $d$ zero eigenvalues in $\mu_{1}',\mu_{2}',...,\mu_{Nd}'.$ Recall
that the eigenvalues of $P$ are $\lambda_{1},\lambda_{2},...,\lambda_{d}$,
each of multiplicity $N$. Denote $\overline{\lambda'}=\{\lambda_{1}'=\lambda_{2}'=...=\lambda_{N}'\leqslant...\leqslant\lambda_{Nd-N}^{'}=\lambda_{Nd-N+1}^{'}=...=\lambda_{Nd}^{'}\}$
where each $\lambda'$ can be either positive or negative, Lemma \ref{lem:1}
suggests that for each $\mu_{v}'$, there exist a $\lambda_{v}'$
of the same order and a $\Theta_{v}$ that has $0\leqslant\Theta_{v}\leqslant\upsilon_{max}$,
such that 
\begin{equation}
\mu_{v}'=\Theta_{v}\lambda_{v}'.\label{eq:eig_J}
\end{equation}
Also, since $\lambda'\neq0$, there are $d$ times that $\Theta_{v}$
takes the value zero. We then conclude that $\Theta_{v}=0$ only happens
when (i) $\left|\lambda_{v}'\right|$ is the smallest, if all those
in $\overline{\lambda'}$ have the same sign, or when (ii) $\lambda_{v}'$
is the smallest positive eigenvalue or the largest negative eigenvalue,
if $\overline{\lambda'}$ has both positive and negative elements.
Since we are dealing with a large population, it is reasonable to
assume that $N\gg d$. Consider (i) where $\lambda_{1}'>0$. If, say,
$\lambda_{N+1}^{'}>\lambda_{1}'$ and $\Theta_{N+1}=0$, then $\mu_{N+1}^{'}=0$.
Since there exists $j\in\{1,2,...,N\}$ where $\Theta_{j}\neq0,\lambda_{j}=\lambda_{1}>0$,
there is $\mu_{N+1}^{'}<\mu_{j}'$ which is a contradiction. The same
applies when $\lambda'$ are all negative and when they can be both.
Equation \eqref{eq:eig_J} then captures the full spectrum of the
Jacobian at the coherent equilibrium.

For the nonzero eigenvalues of $QP$ and $PQ$, since $\Theta_{v}>0$,
the sign of $\mu_{v}'$ is completely determined by the sign of $\lambda_{v}'$,
which is just the eigenvalue of $A$. Then the fully coherent solution
is linearly stable if and only if $A$ is positive definite. In contrast,
should the coupling matrix had any negative eigenvalue, the system
would have at least one dimension that could not synchronize. The
experiment goes further to exhibit that this partly incoherent state
is actually a fully incoherent solution for some of the dimensions,
but its opposite for the rest.

It is then readily inferred that if the system exhibits $\sigma_{i_{1}}\sigma_{i_{2}}\cdots\sigma_{i_{\tau}}\sigma_{i_{\tau+1}}\sigma_{i_{\tau+2}}\cdots\sigma_{i_{d}}\rightarrow\underset{\tau}{\underbrace{\begin{array}{cccc}
0 & 0 & \cdots & 0\end{array}}}\underset{d-\tau}{\underbrace{\begin{array}{cccc}
1 & 1 & \cdots & 1\end{array}}}$ at the end of the transition $A=kA_{initial}$, then the primary
submatrix consisting dimensions $\{i_{\tau+1},i_{\tau+2},...,i_{d}\}$
of $A$ must be positive definite. Because the solution $\underset{\tau}{\underbrace{\begin{array}{cccc}
0 & 0 & \cdots & 0\end{array}}}\underset{d-\tau}{\underbrace{\begin{array}{cccc}
1 & 1 & \cdots & 1\end{array}}}$ is essentially about the fully coherent dimensions $\{i_{\tau+1},i_{\tau+2},...,i_{d}\}$
that decouple from the fully incoherent ones $\{i_{1},i_{2},...,i_{\tau}\}$,
we can set $\sigma_{i_{1}},\sigma_{i_{2}},\cdots,\sigma_{i_{\tau}}$
as zero and analyze the reduced system from \eqref{eq:model-vec}
where $\theta_{i}$ is also rid of the $\{i_{1},i_{2},...,i_{\tau}\}$
dimensions. If the primary submatrix is not positive definite, the
fully coherent solution on the reduced system will in turn be unstable,
then mode $\underset{\tau}{\underbrace{\begin{array}{cccc}
0 & 0 & \cdots & 0\end{array}}}\underset{d-\tau}{\underbrace{\begin{array}{cccc}
1 & 1 & \cdots & 1\end{array}}}$ is unstable on the original system. 

\subsection*{The coherent branch and the onset of synchronization.}

Despite the ultimate system modes we can now relate to the properties
of the coupling matrix, it remains a challenge to analytically derive
their phase diagrams where one or more dimensions may tend to synchronization.

\textcolor{black}{In o}ne of our experiments that is not shown here,
it was observed that with $A=kA_{initial}$, the transition in dimension
$\{\theta_{i1}\}$ or $\{\theta_{i2}\}$ experiences several minor
jumps at lower values of $k$ for even populations as large as $N=1000$,
before gaining coherence consistently with ea\textcolor{black}{ch
increased $k$. }Also, due to the limit size effect, one needs to
tune $k$ slightly below zero to see the onset of synchronization.
Yet for $N=5000$, the transitions of $\sigma_{1,2}$ are smooth enough
to show prospect of fitting analytical results. These transitions
are monotone, have a clear bifurcation at a critical $k$ above zero,
and are in all comparable to that of the second-order. For this reason,
we seek to tackle the $\sigma(A)$ solution in the continuum limit
$N\rightarrow\infty$, where the order parameters are in turn 
\begin{equation}
\rho_{r}(t)=\int_{-\pi}^{\pi}e^{i\theta_{r}(t)}n_{r}(\theta_{r},t)d\theta_{r},\label{eq:rho_continuum}
\end{equation}
$r\in\{1,...,d\}$. The distribution function $n_{r}(\theta_{r})$
can be broken into two parts depending on if its represented population
are fixed on the $r$-th dimension; given that the natural frequencies
are identical on every dimension, we write
\begin{align}
n_{r}(\theta_{r},t) & =n_{rs}(\theta_{r},t)+n_{rd}(\theta_{r},t)\nonumber \\
 & =\int_{-\pi}^{\pi}\cdots\int_{-\pi}^{\pi}\int_{-\infty}^{+\infty}f(\theta_{1},...,\theta_{d},\omega,t)g(\omega)d\omega d\theta_{1}\cdots d\theta_{r-1}d\theta_{r+1}\cdots d\theta_{d}.\label{eq:distribution}
\end{align}

To specify the distribution of the oscillators on each dimension,
we now employ a self-consistent analysis where it is assumed that
the order parameters $\rho_{r}=\sigma_{r}e^{i\Psi_{r}},r\in\{1,...,d\}$
are fixed, therefore the drifting oscillators \textendash{} oscillator
without a fixed point on all of its dimensions \textendash{} must
form a stationary distribution in the space $[0,2\pi]^{d}$. We now
have a density function $f(\theta_{1},...,\theta_{d},\omega)=f(\theta,\omega)$
that is independent of time. Also assume that $\Psi_{r}$ coincides
with the symmetry center of $g(\omega)$, this gives $\rho_{r}=\sigma_{r}$,
and 
\begin{equation}
\dot{\theta}=\omega{\bf 1}_{d}-A\Sigma{\bf sin}(\theta).\label{eq:model-assumption}
\end{equation}

\noindent From here, it is essential to separate the treatment of
each system mode where there may be $\sigma_{r}=0$, because it affects
our way of obtaining the solution of $\dot{\theta}_{r}=0$ from \eqref{eq:model-assumption}
and our interpretation of the solution. Now we have divided the population
into two categories, the fully synchronized oscillators and the drifting
ones, then system modes with $\sigma_{r}=0$ automatically excludes
the existence of the former. But the experiments producing results
like $\sigma_{1}=0,\sigma_{2}=1$ validate that, the oscillators being
drifting does not mean they have zero contribution to any of the order
parameters, i.e., they are impossible to form certain degree of coherence
on any dimension. Moreover, as we shall see, even for system modes
as $\sigma_{1}=1,\sigma_{2}=1$, the drifting oscillators demand further
division to yield an exact theoretic prediction. We will first give
the calculation of the order parameters on the two dimensional population,
then discuss our treatment on arbitrary dimensions.

When one of the order parameters remains zero due to the coupling
matrix and the subsequent instability of the coherent branch, the
other dimension in fact degenerates to the classic Kuramoto model,
e.g., when $\sigma_{2}=0$, \eqref{eq:4} turns into 
\[
\dot{\theta}_{i1}=\omega_{i1}-a_{11}\sigma_{1}\sin(\theta_{i1}-\Psi_{1}),
\]
except the classic coupling strength $K$ is now substituted by the
matrix element $a_{11}$. There is then naturally 
\begin{equation}
1=a_{11}\int_{-\frac{\pi}{2}}^{\frac{\pi}{2}}\cos^{2}\Delta g(a_{11}\sigma_{1}\sin\Delta)d\Delta\label{eq:sigma_classic}
\end{equation}
as the expression of the coherent branch of $\sigma_{1}$ \citep{strogatz2000kuramoto}.
As we scale up $A_{2}$ with $k>0$, $a_{11}$ is strengthened in
the manner of the classic $K$, thus it is not hard to expect that
the transition the order parameter goes through conforms with the
classic case. This simple example also suggests that for population
with higher dimensions, we need to leave out the dimensions with $\sigma_{r}=0$
for a given system mode, and analyze the reduced model 
\begin{equation}
\dot{\theta}'=\omega{\bf 1}_{d'}-A'\Sigma'{\bf sin}(\theta')\label{eq:model_reduced}
\end{equation}
which, compared with \eqref{eq:model-assumption}, involves an $A'$
and a $\Sigma'$ of dimension $d'<d$, and $A'$ is obtained by taking
out the corresponding rows and columns of $A$ to $\sigma_{r}=0$.
The analysis of the reduced system can then build much on the following
calculation of mode 11.

\noindent During the transitions of $\sigma_{1},\sigma_{2}$ from
0 to 1, there are maximally three types of oscillators with different
contributions to the order parameter \textendash{} the fully synchronized,
the orbiting, and the fully incoherent oscillators. For the first
category, consider the model \eqref{eq:model-assumption} and solve
for $\dot{\theta}_{r}=0$ for $r\in\{1,2\}$ simultaneously, we would
have on each dimension a synchronization domain, defined as
\begin{equation}
D_{r}(\omega)=\left\{ \omega:\left|\omega\right|\leqslant\frac{\sigma_{r}}{\left|\left(A^{-1}{\bf 1}_{2}\right)_{r}\right|}\right\} 
\end{equation}
which is always symmetric about the origin despite the value of $r$.
But for the oscillators to have fixed points on both dimensions, the
natural frequency must satisfy 
\[
\omega\in D=\left\{ \omega:\left|\omega\right|<\left[\frac{\sigma_{r}}{\left|\left(A^{-1}{\bf 1}_{2}\right)_{r}\right|}\right]_{\min}=\gamma',r=1,2\right\} .
\]

\noindent In fact, given that $\dot{\theta}_{r}=\omega-\sum_{j=1}^{N}a_{rj}\sigma_{j}\sin(\theta_{j})$
where $\sigma_{j}\neq0$, $\theta$ can either be stationary on all
of its dimensions or none of its dimensions, and any oscillator with
$\omega\in\mathbb{R}\backslash D$ is considered a drifting oscillator.
For $\omega\in D$, we can now derive the fixed point on each dimension,
\begin{equation}
\theta_{r}^{*}=\arcsin\left(\frac{\omega}{\sigma_{r}}\left(A^{-1}{\bf 1}_{2}\right)_{r}\right).\label{eq:r_fixed}
\end{equation}
Since for the entrained population, 
\begin{equation}
f(\theta,\omega)=\prod_{r=1}^{2}\delta\left(\theta_{r}-\theta_{r}^{*}\right),\label{eq:f_fixed}
\end{equation}
put \eqref{eq:distribution}, \eqref{eq:r_fixed}, and \eqref{eq:f_fixed}
in \eqref{eq:rho_continuum}, there is
\begin{align}
\sigma_{r}^{fs} & =\int_{\omega\in D}e^{i\theta_{r}^{*}}g(\omega)d\omega\nonumber \\
 & =\int_{-\gamma'}^{\gamma'}\cos\left(\arcsin\left(\frac{\omega}{\sigma_{r}}\left(A^{-1}{\bf 1}_{2}\right)_{r}\right)\right)g(\omega)d\omega\label{eq:sigma_fs}\\
 & =\frac{\sigma_{r}}{\left(A^{-1}{\bf 1}_{2}\right)_{r}}\int_{-\gamma}^{\gamma}\cos^{2}\Delta_{r}g\left(\frac{\sigma_{r}\sin\Delta_{r}}{\left(A^{-1}{\bf 1}_{2}\right)_{r}}\right)d\Delta_{r},\nonumber 
\end{align}
where $\gamma=\arcsin\left(\frac{\gamma'\left(A^{-1}{\bf 1}_{2}\right)_{r}}{\sigma_{r}}\right),\sin\Delta_{r}=\frac{\omega}{\sigma_{r}}\left(A^{-1}{\bf 1}_{2}\right)_{r}$. 

For population with $\omega\in\mathbb{R}\backslash D$, the natural
frequency exceeds at least one of the synchronization domains $D_{r}$,
and under all circumstances, the oscillators are relatively in motion
to the mean field. However, it does not imply that there is no solution
for $\dot{\theta}_{r}=0$. Here we propose another domain which we
refer to as the original domain on the $r$-th dimension, defined
as 
\[
D_{r}^{o}=\left\{ \omega:\left|\omega\right|\leqslant\sum_{j=1}^{2}\sigma_{j}\left|a_{rj}\right|\right\} .
\]
The original domain depicts the most general condition where $\dot{\theta}_{r}=0$
has a solution. Then, by definition, there is 
\[
D\subseteq D_{r}\subseteq D_{r}^{o}.
\]
We also denote 
\[
D^{o}=\bigcup_{r=1}^{2}D_{r}^{o}.
\]
The contribution of the drifting population to $\sigma_{r}$ distinguishes
between $D^{o}\backslash D$ and $\mathbb{R}\backslash D^{o}$, where
we call the former orbiting oscillators, because when projected onto
the unit sphere, they are constantly drifting on a circle along the
$z$-axis that rotates azimuthally for a small angle after each period
(when $N$ is relatively small, as $N=1000$, and the $\theta_{1}$
mean field is not stationary). We expect this circle, or the orbit,
to be slightly oscillatory about a fixed plane when $N\rightarrow\infty$.
Oscillators with $\omega\in\mathbb{R}\backslash D^{o}$ constitute
the fully incoherent population, we first treat this simpler case
where the continuity equation \eqref{eq:F3} needs to be evaluated
with $\partial f/\partial t=0$, this requires 
\begin{equation}
f(\theta,\omega)=\frac{C(\omega,A,\Sigma)}{\left|\omega{\bf 1}_{d}-A\Sigma{\bf sin}(\theta)\right|_{2}},
\end{equation}
where $C(\omega,A,\Sigma)$ is the normalization constant that guarantees
\[
\int_{-\pi}^{\pi}\int_{-\pi}^{\pi}f(\theta,\omega)d\theta_{1}d\theta_{2}=1
\]
for every $\omega\in\mathbb{R}\backslash D^{o}$. Because each $D_{r}^{o}$
is symmetric about the origin, so is $\mathbb{R}\backslash D^{o}$,
the constant 
\[
\frac{1}{C(\omega,A,\Sigma)}=\int_{-\pi}^{\pi}\int_{-\pi}^{\pi}\frac{1}{\left|\omega{\bf 1}_{2}-A\Sigma{\bf sin}(\theta)\right|_{2}}d\theta_{1}d\theta_{2}
\]
should be invariant against $\omega\rightarrow-\omega$. We can then
write $n_{r}^{fi}(\theta_{r})$ and plug it into \eqref{eq:rho_continuum}
so that the contribution of the fully incoherent oscillators is 
\begin{equation}
\sigma_{r}^{fi}=\int_{-\pi}^{\pi}e^{i\theta_{r}}\int_{-\pi}^{\pi}\int_{\mathbb{R}\backslash D^{o}}\frac{C(\omega,A,\Sigma)g(\omega)}{\left|\omega{\bf 1}_{2}-A\Sigma{\bf sin}(\theta)\right|_{2}}d\omega d\theta_{1}d\theta_{2}.\label{eq:sigma_rd}
\end{equation}
Without loss of generality, for $r=1$, substituting $\theta_{1}\rightarrow\theta_{1}'=\theta_{1}+\pi$,
\eqref{eq:sigma_rd} becomes
\begin{align}
\sigma_{1}^{fi} & =\int_{0}^{\pi}\left[\int_{-\pi}^{\pi}\int_{-\infty}^{+\infty}\frac{e^{i\theta_{1}}C(\omega,A,\Sigma)g(\omega)}{\left|\omega{\bf 1}_{2}-A\Sigma{\bf sin}(\theta)\right|_{2}}d\omega d\theta_{2}\right]d\theta_{1}\nonumber \\
 & -\int_{0}^{\pi}\left[\int_{-\pi}^{\pi}\int_{-\infty}^{+\infty}\frac{e^{i\theta_{1}'}C(\omega,A,\Sigma)g(\omega)}{\left|\omega{\bf 1}_{2}-A\Sigma{\bf sin}(\theta')\right|_{2}}d\omega d\theta_{2}\right]d\theta_{1}'\label{eq:sigma_rd_1}
\end{align}
where ${\bf sin}(\theta')=\left[\begin{array}{cc}
-\sin(\theta_{1}') & \sin(\theta_{2})\end{array}\right]^{T}$. Notice how under the 2-norm of the denominator, each transformation
$\theta_{2}\rightarrow-\theta_{2}$ or $\omega\rightarrow-\omega$
in the first term is a change of the sign before $\sin(\theta_{2})$
or $\omega$, and this arrangement can be found in the second term.
With $C(\omega,A,\Sigma)g(\omega)$ being invariant to $\omega\rightarrow-\omega$,
the two terms in \eqref{eq:sigma_rd_1} are equal, and the contribution
of the fully incoherent oscillators to $\sigma_{r}$ is actually zero.

\begin{figure}
\begin{centering}
\includegraphics[viewport=0bp 0bp 560bp 420bp,width=12cm]{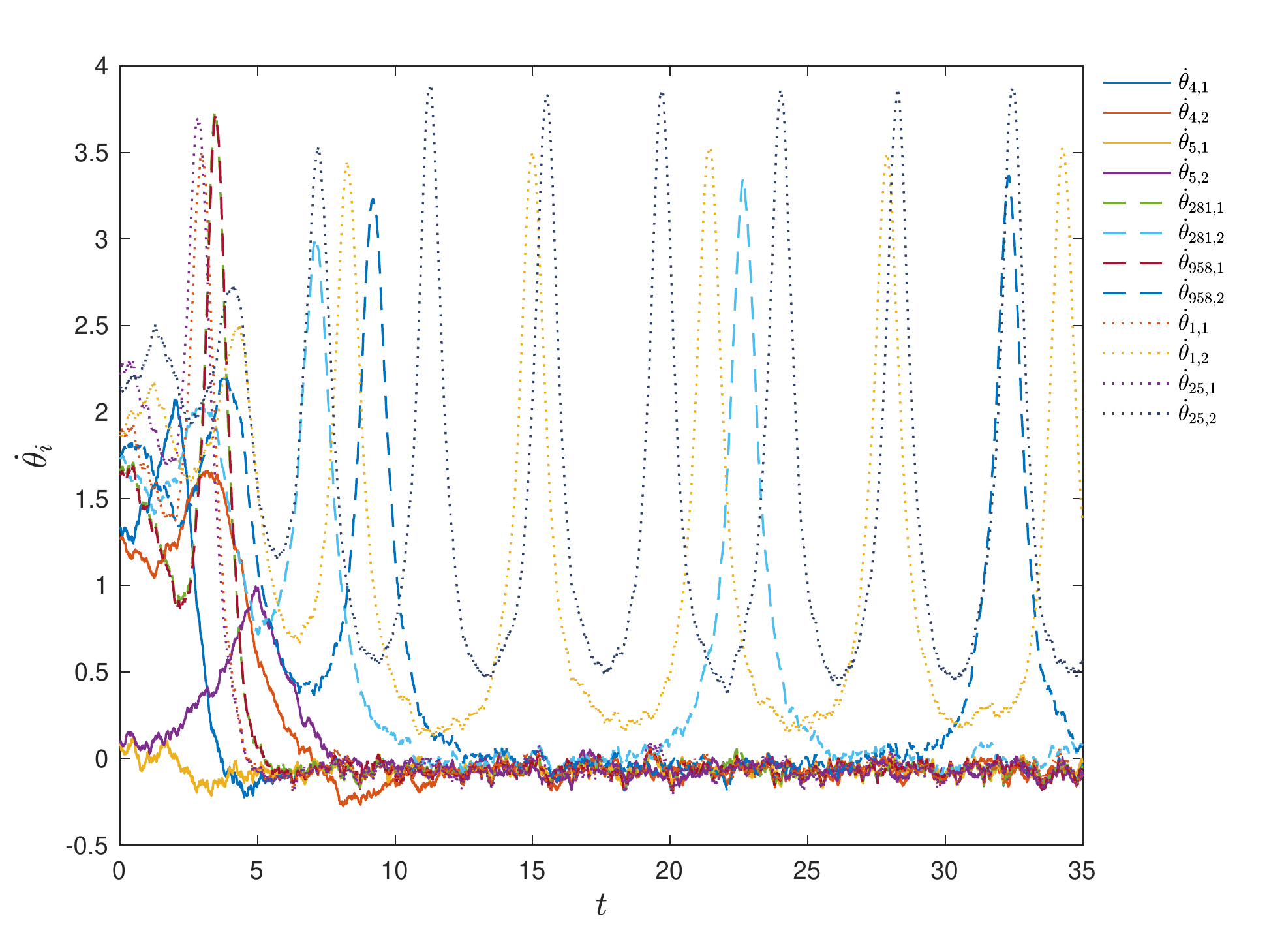}
\par\end{centering}
\caption{\emph{Velocities of sampled oscillators over time}. Six oscillators
are randomly chosen from a population of $N=1000$ at $A=kA_{1},k=0.9$,
when $D_{2}=[-1.6719,1.6719]$, $D_{2}^{o}\backslash D_{2}=[-1.6990,-1.6719)\cup(1.6719,1.6990]$,
and $D_{1}^{o}\backslash D_{2}^{o}=[-2.7245,-1.6990)\cup(1.6990,2.7245]$.
Solid, dashed, and dotted lines are used for oscillators with $\bar{\omega}\in D_{2}$,
$\bar{\omega}\in D_{2}^{o}\backslash D_{2}$, and $\bar{\omega}\in D_{1}^{o}\backslash D_{2}^{o}$,
respectively. Notice that those within the synchronization domain
$D=D_{2}$ have both $\dot{\theta}_{i1}$ and $\dot{\theta}_{i2}$
oscillating about zero with small amplitudes. For $\theta_{281}$
and $\theta_{958}$ whose $\bar{\omega}\in D_{2}^{o}\backslash D_{2}$,
the second dimension is clearly not synchronized, in contrast to the
first dimension. Samples with $\bar{\omega}\in D_{1}^{o}\backslash D_{2}^{o}$
exhibit similar behaviors, except their desynchronized $\dot{\theta}_{i2}$
never really coincide with that of the entrained population. $\theta_{281}$
and $\theta_{958}$, however, experience a bottleneck of the $\dot{\theta}_{i2}$
velocity when they pass through the synchronized population as if
they are themselves entrained, but will eventually move on.}

\label{fig:vel}
\end{figure}

Since to deal with the orbiting oscillators, the relationship between
different domains needs to be specific, we use our experiment with
initial matrix $A_{1}$ as an example, where the data shows $D=D_{2}\subset D_{2}^{o}\subset D_{1}^{o}=D^{o}$
for $k\in[0.2,3.2]$. For the orbiting oscillators whose frequencies
are $\omega\in D^{o}\backslash D$, $\dot{\theta}_{1}=0$ and $\dot{\theta}_{2}=0$
are not solvable simultaneously. The interval $D^{o}\backslash D$
is further divided into $D_{2}^{o}\backslash D$ and $D^{o}\backslash D_{2}^{o}$,
where we each sampled two oscillators and tracked their velocities
over time, as displayed in FIG. \ref{fig:vel}. We did this because
ostensibly, when $\omega\in D_{2}^{o}\backslash D$, both $\dot{\theta}_{1}=0$
and $\dot{\theta}_{2}=0$ are viable, but then it is undecided which
of the two dimensions should be fixed for such oscillators. Experiment
shows in FIG. \ref{fig:vel} that in this case, there is $\dot{\theta}_{1}=0$
instead of $\dot{\theta}_{2}=0$. This is sensible because $\omega$
is well within the margin $D_{1}^{o}$, and also, if there are more
than two dimensions to the oscillators, it is not reasonable to randomly
pick one from $\dot{\theta}_{1}=0,...,\dot{\theta}_{d}=0$ to be unsolvable.
For the rest $\omega\in D^{o}\backslash D_{2}^{o}$, there is obviously
$\dot{\theta}_{1}=0,\dot{\theta}_{2}\neq0$. Based on the above analysis,
\[
f(\theta,\omega)=\frac{C(\omega,A,\Sigma)\delta(\theta_{1}-\theta_{1}^{*})}{\left|\dot{\theta}_{2}\right|},\omega\in D^{o}\backslash D,
\]
where $\theta_{1}^{*}=\arcsin\left(\frac{1}{a_{11}\sigma_{1}}(\omega-a_{12}\sigma_{2}\sin\theta_{2})\right)$
and $C(\omega,A,\Sigma)$ is defined correspondingly. Though $\theta_{1}^{*}$
is not stationary, it guarantees that $\dot{\theta}_{1}$ is close
to zero (as is observed from the experiment), and it allows $\theta_{1}$
to oscillate about $\arcsin\left(\frac{\omega}{a_{11}\sigma_{1}}\right)$,
for which $\theta_{1}$ is stationary when averaged over time. Therefore
the contribution of the orbiting oscillators to $\sigma_{1}$ is 
\begin{align}
\sigma_{1}^{o} & =\int_{-\pi}^{\pi}e^{i\theta_{1}}\int_{-\pi}^{\pi}\int_{D^{o}\backslash D}\frac{C(\omega,A,\Sigma)\delta(\theta_{1}-\theta_{1}^{*})g(\omega)}{\left|\dot{\theta}_{2}\right|}d\omega d\theta_{2}d\theta_{1}\nonumber \\
 & =\int_{-\pi}^{\pi}\int_{D^{o}\backslash D}\frac{C(\omega,A,\Sigma)e^{i\theta_{1}^{*}}g(\omega)}{\left|\left(1-\frac{a_{21}}{a_{11}}\right)\omega-\frac{|A|}{a_{11}}\sigma_{2}\sin\theta_{2}\right|}d\omega d\theta_{2}\label{eq:sigma1o}\\
 & =\int_{-\pi}^{\pi}\int_{D^{o}\backslash D}\frac{C(\omega,A,\Sigma)g(\omega)\cos\left(\arcsin\left(\frac{\omega}{a_{11}\sigma_{1}}-\frac{a_{12}\sigma_{2}\sin\theta_{2}}{a_{11}\sigma_{1}}\right)\right)}{\left|\left(1-\frac{a_{21}}{a_{11}}\right)\omega-\frac{|A|}{a_{11}}\sigma_{2}\sin\theta_{2}\right|}d\omega d\theta_{2},\nonumber 
\end{align}
where 
\[
C(\omega,A,\Sigma)=\int_{-\pi}^{\pi}\frac{1}{\left|\left(1-\frac{a_{21}}{a_{11}}\right)\omega-\frac{|A|}{a_{11}}\sigma_{2}\sin\theta_{2}\right|}d\theta_{2},
\]
and the last line of \eqref{eq:sigma1o} stands because the imaginary
part 
\[
\int_{-\pi}^{\pi}\int_{D^{o}\backslash D}\frac{C(\omega,A,\Sigma)g(\omega)\left(\frac{\omega}{a_{11}\sigma_{1}}-\frac{a_{12}\sigma_{2}\sin\theta_{2}}{a_{11}\sigma_{1}}\right)}{\left|\left(1-\frac{a_{21}}{a_{11}}\right)\omega-\frac{|A|}{a_{11}}\sigma_{2}\sin\theta_{2}\right|}d\omega d\theta_{2}=0.
\]
Combining our analysis of \eqref{eq:sigma_fs}, \eqref{eq:sigma_rd}
and \eqref{eq:sigma1o}, the order parameters for mode 11 are derived
as
\begin{align}
\sigma_{1} & =\int_{-\gamma'}^{\gamma'}\cos\left(\arcsin\left(\frac{\omega}{\sigma_{1}}\left(A^{-1}{\bf 1}_{2}\right)_{1}\right)\right)g(\omega)d\omega\label{eq:sigma_12}\\
 & +\int_{-\pi}^{\pi}\int_{D^{o}\backslash D}\frac{C(\omega,A,\Sigma)g(\omega)\cos\left(\arcsin\left(\frac{\omega}{a_{11}\sigma_{1}}-\frac{a_{12}\sigma_{2}\sin\theta_{2}}{a_{11}\sigma_{1}}\right)\right)}{\left|\left(1-\frac{a_{21}}{a_{11}}\right)\omega-\frac{|A|}{a_{11}}\sigma_{2}\sin\theta_{2}\right|}d\omega d\theta_{2},\nonumber \\
\sigma_{2} & =\int_{-\gamma'}^{\gamma'}\cos\left(\arcsin\left(\frac{\omega}{\sigma_{2}}\left(A^{-1}{\bf 1}_{2}\right)_{2}\right)\right)g(\omega)d\omega.\nonumber 
\end{align}
In fact, for the $d$ dimensional population, the calculation of the
synchronization branches follows similar procedures. We mentioned
that the dimensions with $\sigma_{r}=0$ must be left out for a given
system mode, and the reduced system \eqref{eq:model_reduced} is considered
instead. Therefore, without loss of generality, we assume $d'=d$,
$D=D_{1}\subsetneqq D_{2}^{o}\subsetneqq...\subsetneqq D_{d}^{o}$,
and look for $\sigma_{r}$'s expression in mode $\left(11\cdots1\right)_{d}$.
When the natural frequency lands in the interval $D_{r'}^{o}\backslash D_{r'-1}^{o},r\geqslant3$
or $D_{r'}^{o}\backslash D_{r'-1},r=2$, the dimensions automatically
divide into $\{r',r'+1,...,d\}$ where $\dot{\theta}_{r\in\{r',...,d\}}=0$
is solvable, and $\{1,...,r'-1\}$ otherwise. One can write the solutions
to $\dot{\theta}_{r\in\{r',...,d\}}=0$ as the function of $\theta_{1},...,\theta_{r'-1}$,
such that 
\begin{equation}
\theta_{r}^{*}=F_{r}\left(\theta_{1},\theta_{2},...,\theta_{r'-1}\right),\label{eq:fix_highdim}
\end{equation}
where $F_{r}$ is obtained by solving
\[
\bar{A}_{r'd}\begin{bmatrix}\sigma_{r'}\\
 & \ddots\\
 &  & \sigma_{d}
\end{bmatrix}\begin{bmatrix}\sin\theta_{r'}\\
\vdots\\
\sin\theta_{d}
\end{bmatrix}=\bar{\omega}{\bf 1}_{d-r'+1}-\begin{bmatrix}\sum_{j=1}^{r'-1}a_{r'j}\sigma_{j}\sin\theta_{j}\\
\vdots\\
\sum_{j=1}^{d-1}a_{dj}\sigma_{j}\sin\theta_{j}
\end{bmatrix}.
\]
$\bar{A}_{r'd}$ is the principle sub matrix of $A$ containing the
$r'$-th to the $d$-th dimension. Since $A$ is positive definite,
all of its principle minors are positive, which means $\bar{A}_{r'd}$
should be invertible, and therefore, the expression in \eqref{eq:fix_highdim}
is uniquely obtained. All fixed points are essentially the linear
combination of $\bar{\omega},\sin\theta_{1},\sin\theta_{2},...,\sin\theta_{r'-1}$.
We then write the density function for $\omega\in D_{r'}^{o}\backslash D_{r'-1}^{o}$
as 
\begin{equation}
f_{r'-1}^{r'}=\frac{C(\omega)\prod_{r=r'}^{d}\delta(\theta_{r}-\theta_{r}^{*})}{\left(\sum_{r=1}^{r'-1}\dot{\theta}_{r}^{2}\left(\omega,\theta_{1},...,\theta_{r'-1}\right)\right)^{1/2}}
\end{equation}
where the velocity $\dot{\theta}_{r}$ is also a linear combination
of $\bar{\omega},\sin\theta_{1},\sin\theta_{2},...,\sin\theta_{r'-1}$.
Then the contribution of $\omega\in D_{r'}^{o}\backslash D_{r'-1}^{o}$
for a specific $\sigma_{\bar{r}}$ is, if $\bar{r}\leqslant r'-1$,
\[
\sigma_{\bar{r}(D_{r'}^{o}\backslash D_{r'-1}^{o})}^{o}=\int_{-\pi}^{\pi}d\theta_{1}\cdots\int_{-\pi}^{\pi}d\theta_{r'-1}\int_{D_{r'}^{o}\backslash D_{r'-1}^{o}}d\omega\frac{C(\omega)e^{i\theta_{\bar{r}}}g(\omega)}{\left(\sum_{r=1}^{r'-1}\dot{\theta}_{r}^{2}\left(\omega,\theta_{1},...,\theta_{r'-1}\right)\right)^{1/2}}=0,
\]
for the same reason as \eqref{eq:sigma_rd_1}. This means $\sigma_{\bar{r}}^{o}$
should only take into account frequency intervals that allow the oscillators
to be fixed on dimension $\bar{r}$. And for $\bar{r}>r'-1$, the
contribution on the above interval is
\begin{align}
\sigma_{\bar{r}(D_{r'}^{o}\backslash D_{r'-1}^{o})}^{o} & =\int_{-\pi}^{\pi}d\theta_{1}\cdots\int_{-\pi}^{\pi}d\theta_{r'-1}\int_{D_{r'}^{o}\backslash D_{r'-1}^{o}}d\omega\frac{C(\omega)e^{i\theta_{\bar{r}^{*}}}g(\omega)}{\left(\sum_{r=1}^{r'-1}\dot{\theta}_{r}^{2}\left(\omega,\theta_{1},...,\theta_{r'-1}\right)\right)^{1/2}}\nonumber \\
 & =\int_{-\pi}^{\pi}d\theta_{1}\cdots\int_{-\pi}^{\pi}d\theta_{r'-1}\int_{D_{r'}^{o}\backslash D_{r'-1}^{o}}d\omega\frac{C(\omega)g(\omega)\cos\left(F_{\bar{r}}\left(\theta_{1},...,\theta_{r'-1}\right)\right)}{\left(\sum_{r=1}^{r'-1}\dot{\theta}_{r}^{2}\left(\omega,\theta_{1},...,\theta_{r'-1}\right)\right)^{1/2}}.\label{eq:sigma_rbar}
\end{align}
Summing \eqref{eq:sigma_rbar} over $r'\leqslant\bar{r}$, we arrive
at the expression of $\sigma_{\bar{r}}$ in terms of the orbiting
oscillators $\sigma_{\bar{r}}^{o}$. Combining $\sigma_{\bar{r}}^{o}$
with $\sigma_{\bar{r}}^{fi}=0$ and 
\[
\sigma_{\bar{r}}^{fs}=\int_{-\gamma'}^{\gamma'}\cos\left(\arcsin\left(\frac{\omega}{\sigma_{\bar{r}}}\left(A^{-1}{\bf 1}_{d}\right)_{\bar{r}}\right)\right)g(\omega)d\omega,
\]
we obtain the full equation of the synchronization branch of $\sigma_{\bar{r}}$
with arbitrary dimensional population.

\section*{Discussion}

In this paper, we introduced the matrix-coupling mechanism into the
study of the Kuramoto model, which gives rise to a novel macroscopic
phenomenon where any combination of coherence/incoherence of the dimensions
of the population is possible. Usage of the term ``coupling matrix''
has been pervasive in previous literatures in network science \citep{li2008synchronization,patra2016statistics,Coss2018},
or even sometimes referring to the connectivity matrix in brain theory
\citep{lameu2018alterations}, but caution is due where it stands
for an adjacency matrix as the cases above that just arranges the
scalar-valued pairwise coupling strengths into a matricial form. An
explicit matrix effect on the diffusive coupling between high-dimensional
Kuramoto oscillators was rarely studied save for a few exceptions
\citep{Buzanello2022}, not to say that in the representation of equation
\eqref{eq:model}, to the best knowledge of the authors.

Our main contributions are as follows: (1) for a $d$ dimensional
population, we proved the existence of $2^{d}$ system modes under
matrix coupling where each dimension of the population can be fully
coherent or fully incoherent; (2) necessary and/or sufficient conditions
are established between the ``positiveness'' of the coupling matrix,
or the lack thereof, and the stable system mode encoded in a combination
of the binary dimensional order parameters; (3) steady-state solutions
of the order parameters are derived for arbitrarily many dimensions
that tend towards synchronization through a second-order phase transition,
along with the complete phase diagram for the two dimensional population;
(4) an analytical estimation of the critical values of the coupling
matrix at which each of the dimensions starts to synchronize is obtained
through a multi-variable Fourier analysis; (5) in the numerical study
elaborated in the Supplementary Material, we demonstrate the existence
of explosive synchronization/desynchronization of the dimensions,
as we show that the system modes are interchangeable through matrix
manipulation. The extensive analysis we performed in this work sheds
light on the occurrence of the binary system modes and their statistical
features, which lays the foundation for further exploration of this
model. The last finding is somewhat exceptional since it was believed
that with Kuramoto-based interaction models, fast switching between
the states must happen when there are degree-correlated natural frequencies
\citep{gomez2011explosive}, a global feedback from the order parameter
\citep{zhang2015explosive,Dai2020DiscontinuousTA}, or higher-order
effects with explicit three-way or even four-way interactions \citep{Skardal2019AbruptDA,Skardal2020HigherOI,Millan2019ExplosiveHK}.
Meanwhile, our model admitted explosive synchronization/desynchronization
without the above mechanisms, which may provide fresh insight into
the understanding of such phenomena that have been related to epileptic
seizures, bistable perception and other behaviors of the brain \citep{andrew1989seizure,wang2017small,wang2013brain}.
Our experiment also suggests a simple way of controlling the macroscopic
state to switch between pairs of system modes which may turn out effective
with further investigations in applications.

The main limitation of this work is that the condition on the principle
submatrix that is linked to a nontrivial combination of the order
parameters, i.e., that which excludes the fully coherent state and
the fully incoherent state, is necessary but not sufficient. The Fourier
analysis we conducted on the fully incoherent solution, though proves
successful in giving analytical predictions of the critical coupling
strength, does not suggest which system mode will emerge from the
developing matrix effect. A future aim is thus a possible variation
of the ansatz proposed in \citep{Ott2008,ott2009long} as in \citep{chandra2019complexity},
that may lead to a complete phase reduction to lower dimensions and
capture all the essential phenomena demonstrated in this work.

The proposed model is distinct from other network dynamics by the
unique macroscopic phenomena it displays. For the majority of the
variations of the Kuramoto model, including the multi-layer \citep{zhang2015explosive},
the higher-order \citep{Millan2019ExplosiveHK,Skardal2020HigherOI},
and the high-dimensional \citep{zhu2013synchronization,chandra2019continuous,Dai2020DiscontinuousTA}
ones, great effort was dedicated to the synchronization phenomenon
and the nuance in the phase transition toward it, which is also a
main focus of this work. But in terms of desynchronization that is
indeed discussed on a high-dimensional population, as we stated about
FIG. \ref{fig:IDPT_sphere}, a direct comparison is difficult to draw
when \citep{Chandra2019ObservingMT,chandra2019continuous} defined
the population on the unit sphere, since the coordinate transformation
is not allowed when all of the components fall in $[0,2\pi]$. If
we loosely apply the idea to a population on the 3D unit sphere that
one angular variable is synchronized while the other is not, in analogy
to mode 10 or mode 01 in this work, then as in ref. \citep{chandra2019continuous},
all of these incoherent states, together with an infinite continuum
of distributions whose centroid is the origin, collapse into a single
representation of a zero-magnitude order parameter. According to the
precessing equation $d\overrightarrow{\sigma}_{i}/dt=W_{i}\overrightarrow{\sigma}_{i}$
in this case and the distribution $G(W_{i})$, the aforementioned
system modes are not steady-state distributions that can be observed
in their system. 

Another idea reminiscent of our discovery is the isolated desynchronization
in cluster synchronization studied on a rather general form of the
complex network \citep{Pecora2014,lodi2021one,Cho2017}, where a population
of high-dimensional oscillators is divided into a number of subsets
that each evolves along a unique, synchronized trajectory. On occasion,
as the mentioned works proved, synchronized solutions may lose stability
for some of these subsets while others carry on, which can be detected
by identifying the symmetries in the network. Although this description
bears certain similarity to our phenomenon, a rigorous comparison
reveals the difference considering the population studied in this
work is also high-dimensional. Thus by their definition, no cluster
is formed for any of the system modes with one of the dimensional
order parameters being zero, as it indicates desynchronization for
the entire population. 

The most fruitful applications of our results may be found, as we
believe, in the field of neuroscience, where a desynchronized state
in part of an ensemble is particularly meaningful. A notable example
is the unihemispheric slow-wave sleep observed in birds and aquatic
mammals like the dolphin \citep{mukhametov1977interhemispheric,Rattenborg2000BehavioralNA},
where the two hemispheres of the brain alternate between the resting
state (synchronized state) and the activated state (desynchronized
state), the two behaviors existing independently at the same time.
The analogy provided by our numerical experiment is that, since the
synchronization and desynchronization of the two dimensions are separated,
their combinations suggest four qualitatively distinct modes of the
system that are possible to switch between one and another through
manipulating the matrix elements, which may facilitate our understanding
of the mentioned phenomenon. On the other hand, in terms of the information
storage and processing of oscillatory neural networks (ONN) \citep{Raychowdhury2019,Csaba2020}
that was introduced based on the Kuramoto model \citep{Hoppensteadt2000},
our finding suggests a new way to reconcile the phase-relation logic
of the ONN with the boolean logic adopted by the traditional von Neumann
machine, where the former encodes information with the in-phase (logic
``0'') or anti-phase (logic ``1'') distributions of the oscillators.
Here we have shown that through a set of statistical measures, the
dimensions as digits naturally carry boolean information.

\section*{Method}

\subsection*{Linear Stability of the Incoherent Solution}

Here we are interested in the stability of the state where the oscillators
are incoherent on all dimensions, i.e., $\sigma_{1}=\sigma_{2}=0$.
We aim to derive an explicit estimation of the critical value of the
coupling matrix from which the system starts the transition towards
coherence on each dimension, under the presumption that such a transition
indeed occurs. Assume that $N\rightarrow\infty$, we consider the
population as a flow on a two-dimensional plane with $2\pi$\textendash periodic
bounds. For our purpose, the flow has a perturbed density function
\begin{equation}
f(\theta_{1},\theta_{2},\omega,t)=\frac{1}{4\pi^{2}}+\epsilon\eta(\theta_{1},\theta_{2},\omega,t),\label{eq:F1}
\end{equation}
where $\epsilon$ is small, and $f(\theta_{1},\theta_{2},\omega,t)d\theta_{1}d\theta_{2}$
denotes the fraction of oscillators with natural frequency $\omega$
on both dimensions, that are found in the infinitesimal area $[\theta_{1},\theta_{1}+d\theta_{1}]\times[\theta_{2},\theta_{2}+d\theta_{2}]$
at time $t$. Therefore the normalization condition requires that
\begin{equation}
\int_{-\pi}^{\pi}\int_{-\pi}^{\pi}\eta(\theta_{1},\theta_{2},\omega,t)d\theta_{1}d\theta_{2}=0.\label{eq:F2}
\end{equation}
Following the method of \citep{strogatz2000kuramoto}, given that
the oscillators are conserved, we set up the continuity equation
\begin{equation}
\partial f/\partial t+\nabla(f(\theta_{1},\theta_{2},\omega,t){\bf v})=0\label{eq:F3}
\end{equation}
where the velocity field is that of equation \eqref{eq:4}. Inserting
\eqref{eq:4} and \eqref{eq:F1} into \eqref{eq:F3} gives
\begin{align}
\epsilon\frac{\partial\eta}{\partial t} & -\left(\frac{1}{4\pi^{2}}+\epsilon\eta\right)\left[a_{11}\sigma_{1}\cos(\theta_{1}-\Psi_{1})+a_{22}\sigma_{2}\cos(\theta_{2}-\Psi_{2})\right]\nonumber \\
 & +\epsilon\left\{ \frac{\partial\eta}{\partial\theta_{1}}\left[\omega-a_{11}\sigma_{1}\sin(\theta_{1}-\Psi_{1})-a_{12}\sigma_{2}\sin(\theta_{2}-\Psi_{2})\right]\right.\label{eq:F4}\\
 & \left.+\frac{\partial\eta}{\partial\theta_{2}}\left[\omega-a_{21}\sigma_{1}\sin(\theta_{1}-\Psi_{1})-a_{22}\sigma_{2}\sin(\theta_{2}-\Psi_{2})\right]\right\} =0.\nonumber 
\end{align}

To perform Fourier analysis, we expand $\eta(\theta_{1},\theta_{2},\omega)$
into Fourier series
\begin{equation}
\eta(\theta_{1},\theta_{2},\omega,t)=\sum_{m=-\infty}^{+\infty}\sum_{n=-\infty}^{+\infty}c_{m,n}(\omega,t)e^{im\theta_{1}}e^{in\theta_{2}}.
\end{equation}
Note that for real-valued functions, $c_{m,n}(\omega,t)=c_{-m,-n}^{*}(\omega,t)$.
As we evaluate the order parameter $\sigma_{1}e^{i\Psi_{1}}$ for
$N\rightarrow\infty$, 
\begin{align*}
\sigma_{1}e^{i\Psi_{1}} & =\epsilon\int_{-\pi}^{\pi}\int_{-\pi}^{\pi}\int_{-\infty}^{+\infty}\eta(\theta_{1},\theta_{2},\omega,t)g(\omega)e^{i\theta_{1}}d\omega d\theta_{1}d\theta_{2}\\
 & =\epsilon\sum_{m=-\infty}^{+\infty}\sum_{n=-\infty}^{+\infty}\int_{-\pi}^{\pi}e^{i(m+1)\theta_{1}}d\theta_{1}\int_{-\pi}^{\pi}e^{in\theta_{2}}d\theta_{2}\int_{-\infty}^{+\infty}c_{mn}(\omega,t)g(\omega)d\omega\\
 & =4\pi^{2}\epsilon\int_{-\infty}^{+\infty}c_{-1,0}(\omega,t)g(\omega)d\omega,
\end{align*}
it is noted that only term $c_{-1,0}(\omega,t)e^{-i\theta_{1}}+c.c.$
contributes, $c.c.$ representing the complex conjugate of the previous
term. This similarly applies to 
\[
\sigma_{2}e^{i\Psi_{2}}=4\pi^{2}\epsilon\int_{-\infty}^{+\infty}c_{0,-1}(\omega,t)g(\omega)d\omega
\]
where harmonics $c_{0,-1}(\omega,t)e^{-i\theta_{2}}+c.c.$ alone contribute
to the order parameter. It is further derived that
\begin{align}
\sigma_{1}\cos(\theta_{1}-\Psi_{1}) & =\text{Re}\left[\sigma_{1}e^{i(\Psi_{1}-\theta_{1})}\right]\nonumber \\
 & =4\pi^{2}\epsilon\text{Re}\left[e^{-i\theta_{1}}\int_{-\infty}^{+\infty}c_{-1,0}(\omega,t)g(\omega)d\omega\right]\nonumber \\
 & =2\pi^{2}\epsilon e^{-i\theta_{1}}\int_{-\infty}^{+\infty}c_{-1,0}(\omega,t)g(\omega)d\omega+c.c.,\label{eq:F6}
\end{align}
and 
\begin{equation}
\sigma_{2}\cos(\theta_{2}-\Psi_{2})=2\pi^{2}\epsilon e^{-i\theta_{2}}\int_{-\infty}^{+\infty}c_{0,-1}(\omega,t)g(\omega)d\omega+c.c.,\label{eq:F7}
\end{equation}
\begin{align}
\sigma_{1}\sin(\theta_{1}-\Psi_{1}) & =-\text{Im}\left[\sigma_{1}e^{i(\Psi_{1}-\theta_{1})}\right]\nonumber \\
 & =-4\pi^{2}\epsilon\text{Im}\left[e^{-i\theta_{1}}\int_{-\infty}^{+\infty}c_{-1,0}(\omega,t)g(\omega)d\omega\right]\nonumber \\
 & =-\frac{1}{i}2\pi^{2}\epsilon\left(e^{-i\theta_{1}}\int_{-\infty}^{+\infty}c_{-1,0}(\omega,t)g(\omega)d\omega-e^{i\theta_{1}}\int_{-\infty}^{+\infty}c_{-1,0}^{*}(\omega,t)g(\omega)d\omega\right)\nonumber \\
 & =2\pi^{2}i\epsilon e^{-i\theta_{1}}\int_{-\infty}^{+\infty}c_{-1,0}(\omega,t)g(\omega)d\omega+c.c.,
\end{align}
\begin{equation}
\sigma_{2}\sin(\theta_{2}-\Psi_{2})=2\pi^{2}i\epsilon e^{-i\theta_{2}}\int_{-\infty}^{+\infty}c_{0,-1}(\omega,t)g(\omega)d\omega+c.c..
\end{equation}
We separate the terms that governs the evolution of $\eta(\theta_{1},\theta_{2},\omega,t)$
in \eqref{eq:F4} from the others,
\begin{align}
\eta(\theta_{1},\theta_{2},\omega,t) & =\eta_{0}+\tilde{\eta}\label{eq:F10}\\
 & =c_{1,0}(\omega,t)e^{i\theta_{1}}+c_{0,1}(\omega,t)e^{i\theta_{2}}+c.c.+\tilde{\eta},\nonumber 
\end{align}
where $\tilde{\eta}$ stands for the rest of the harmonics. Equation
\eqref{eq:F2} implies that $c_{0,0}=0$, therefore $\eta_{0}$ can
be considered the fundamental mode of $\eta$ where the disturbances
on the $\theta_{1}$ dimension and the $\theta_{2}$ dimension are
decoupled. Inserting \eqref{eq:F6}\textendash \eqref{eq:F10} into
\eqref{eq:F4} and linearizing result in
\begin{equation}
\frac{\partial c_{1,0}}{\partial t}e^{i\theta_{1}}+\frac{\partial c_{0,1}}{\partial t}e^{i\theta_{2}}+i\omega\left(c_{1,0}e^{i\theta_{1}}+c_{0,1}e^{i\theta_{2}}\right)-\frac{1}{2}\left(a_{11}\overline{c_{1,0}}e^{i\theta_{1}}+a_{22}\overline{c_{0,1}}e^{i\theta_{2}}\right)=0,\label{eq:F11}
\end{equation}
where $\overline{c(\omega)}=\int_{-\infty}^{+\infty}c(\omega)g(\omega)d\omega$.
Solving \eqref{eq:F11} with respect to $e^{i\theta_{1}}$and $e^{i\theta_{2}}$,
we derive
\begin{equation}
\frac{\partial c_{1,0}(\omega,t)}{\partial t}+i\omega c_{1,0}(\omega,t)-\frac{a_{11}}{2}\int_{-\infty}^{+\infty}c_{1,0}(\omega,t)g(\omega)d\omega=0,\label{eq:F12}
\end{equation}
and 
\begin{equation}
\frac{\partial c_{0,1}(\omega,t)}{\partial t}+i\omega c_{0,1}(\omega,t)-\frac{a_{22}}{2}\int_{-\infty}^{+\infty}c_{0,1}(\omega,t)g(\omega)d\omega=0.\label{eq:F13}
\end{equation}
To determine the stability of solution \eqref{eq:F1}, set $c_{1,0}(\omega,t)=b_{1,0}(\omega)e^{\kappa_{1}t},c_{0,1}(\omega,t)=b_{0,1}(\omega)e^{\kappa_{2}t}$,
it is obvious that \eqref{eq:F12} and \eqref{eq:F13} have the same
discrete spectrum as that of the classic model derived in \citep{Strogatz1991StabilityOI},
which means $\kappa_{1},\kappa_{2}$ are derived from the characteristic
equations 
\begin{align}
1 & =\frac{a_{11}}{2}\int_{-\infty}^{+\infty}\frac{\kappa_{1}}{\kappa_{1}^{2}+\omega^{2}}g(\omega)d\omega,\label{eq:F14}\\
1 & =\frac{a_{22}}{2}\int_{-\infty}^{+\infty}\frac{\kappa_{2}}{\kappa_{2}^{2}+\omega^{2}}g(\omega)d\omega.\label{eq:F15}
\end{align}
In this work we have adopted the standard Lorentzian distribution
with $g(\omega)=\frac{1}{\pi(\omega^{2}+1)}$, this leaves the critical
value of $a_{11}$($a_{22}$, resp.) at which the effect of $\eta(\theta_{1},\theta_{2},\omega,t)$
on the $\theta_{1}$($\theta_{2}$, resp.) dimension renders \eqref{eq:F1}
unstable to be
\begin{equation}
\left[a_{11}\right]_{critical}=\left[a_{22}\right]_{critical}=\frac{2}{\pi g(0)}.
\end{equation}
This result is readily generalized into $d$ dimensional cases, if
we recognize
\[
f(\theta,\omega,t)=\frac{1}{(2\pi)^{d}}+\epsilon\eta(\theta,\omega,t)
\]
and 
\[
\eta(\theta,\omega,t)=\sum_{n_{1}=-\infty}^{+\infty}\cdots\sum_{n_{d}=-\infty}^{+\infty}c_{(n_{1},n_{2},...,n_{d})}(\omega,t)e^{in_{1}\theta_{1}+\cdots+in_{d}\theta_{d}}.
\]
Similarly, one will find that only the first harmonics contribute
to $\eta$ in the sense of \eqref{eq:F3}. Denoting the $d$-tuple,
$\left(\begin{array}{ccccccc}
0 & \cdots & 0 & \underset{r\text{-th}}{1} & 0 & \cdots & 0\end{array}\right)$ as $\mathbb{I}_{r}$, we mention that $c_{\mathbb{I}_{r}}=\frac{1}{2\pi}\int_{-\pi}^{\pi}\eta e^{-in_{r}\theta_{r}}d\theta_{r}$,
therefore $c_{-\mathbb{I}_{r}}=c_{\mathbb{I}_{r}}^{*}$, and the complex
conjugate terms evolve under the same dynamics. Then still, at $O(\epsilon)$,
we have 
\[
\sum_{r=1}^{d}e^{i\theta_{r}}\left[\frac{\partial c_{\mathbb{I}_{r}}(\omega,t)}{\partial t}+i\omega c_{\mathbb{I}_{r}}-\frac{a_{rr}}{2}\int_{-\infty}^{+\infty}c_{\mathbb{I}_{r}}(\omega,t)g(\omega)d\omega\right]=0.
\]

Recall that in the numerical experiment, we scale all the elements
of $A$ by a real number $k$ and increase it slightly at each step.
The coupling matrix $A$ is required to be real symmetric, but the
diagonal elements are not necessarily identical. This means that whichever
is positive and larger will reach its critical value first, and lead
to the phase transition to synchronization on its corresponding dimension.
Equations \eqref{eq:F14} and \eqref{eq:F15} also suggest that if
a diagonal element $a_{rr}$ is negative, the characteristic equation
on $\kappa_{r}$ will be unsolvable for a discrete spectrum, leaving
only the continuous spectrum on the imaginary line \citep{strogatz2000kuramoto},
which means $\sigma_{r}=0$ remains neutrally stable against the disturbance
from $\eta$ on the $r$-th dimension for the entire time. As a result,
when $a_{rr}<0$, the $r$-th dimension of the population will not
see a transition toward synchronization with an increasing positive
$k$.

\subsection*{Numerical Simulation.}

The four initial matrices adopted to generate FIG. \ref{fig:IDPT_sphere}
are $A_{1}=\begin{bmatrix}4.27 & 0.11\\
0.11 & 3.20
\end{bmatrix},A_{2}=\begin{bmatrix}3.80 & 7.54\\
7.54 & 2.86
\end{bmatrix},A_{3}=\begin{bmatrix}4.20 & 12.5\\
12.5 & 10
\end{bmatrix},A_{4}=\begin{bmatrix}-5.11 & 6.28\\
6.28 & -11.89
\end{bmatrix}$. The differential equation \eqref{eq:model} was numerically integrated
using a 5th order Runge-Kutta formula, with relative error tolerance
$10^{-5}$ and step length $\Delta t=0.01$. The natural frequency
$\omega_{i}=\bar{\omega}_{i}{\bf 1}_{2}$ of the population follows
the standard Lorentzian distribution $g(\omega)=1/[\pi(\omega^{2}+1)]$.
At each value of $k$, we integrated until the order parameter $\sigma_{1}(\sigma_{2})$
had settled around an equilibrium, then took the average of $\sigma(t)$
for the last 1500 integration units. The experiment performed on the
three dimensional population has adopted the same convention.

In our test of the $10^{4}$ initial values, we determine the system
mode a particular $\theta(0)$ is attracted to based on the numerical
result in FIG. \ref{fig:IDPT_sphere} and the following experimental
fact: whenever a dimension of the population is tending to an incoherent
solution from an almost (but not exactly) uniform initial distribution,
be it with $N=1000$ or with $N=5000$, its order parameter $\sigma_{r}$
significantly overcomes the limit size effect which causes $\sigma_{r}=\epsilon>0$
at $k=0$, and becomes almost vanishing for $k>k_{critical}$. For
$N=100$, there is approximately $\epsilon=0.1$. Consequently, we
set up the criterion that if $\sigma_{1}\sigma_{2}$ is comparable
to the data in FIG. 3(a) in the sense that $\sigma_{r}(k)\in[\sigma_{r}^{data_{-}a}(k)-0.1,\sigma_{r}^{data_{-}a}(k)+0.1],r=1,2$,
then the $\theta(0)$ is attracted to mode 11; if $\sigma_{2}(k)<0.1$
and $\sigma_{1}(k)\in[\sigma_{1}^{data_{-}b}(k)-0.1,\sigma_{1}^{data_{-}b}(k)+0.1]$,
we consider $\theta(0)$ to be attracted to mode 10. The criteria
for mode 01 and mode 00 are similarly defined, with $\sigma_{1}(k)<0.1,$$\sigma_{2}(k)\in[\sigma_{2}^{data_{-}c}(k)-0.1,\sigma_{2}^{data_{-}c}(k)+0.1]$
and $\sigma_{1}(k)<0.1,\sigma_{2}(k)<0.1.$ Note that all the initial
value tests are performed on the same population, i.e., the natural
frequencies $\{\omega_{i}\}$ are controlled.

\section*{\textemdash \textemdash \textemdash \textemdash \textemdash \textendash{}}

\bibliographystyle{unsrt}
\addcontentsline{toc}{section}{\refname}\bibliography{42_Users_sheepawe_Library_CloudStorage_Dropbox_Matrix-Kuramoto-3rd_arxiv_lib_K2dim}

\begin{thebibliography}{10}

\bibitem{Boccaletti2018}
Stefano Boccaletti, Alexander~N. Pisarchik, Charo~I. del Genio, and Andreas
  Amann.
\newblock {\em Synchronization: From Coupled Systems to Complex Networks}.
\newblock Cambridge University Press, 2018.

\bibitem{jadbabaie2003coordination}
Ali Jadbabaie, Jie Lin, and A~Stephen Morse.
\newblock Coordination of groups of mobile autonomous agents using nearest
  neighbor rules.
\newblock {\em IEEE Transactions on automatic control}, 48(6):988--1001, 2003.

\bibitem{Ren2005}
Wei Ren, R.W. Beard, and E.M. Atkins.
\newblock A survey of consensus problems in multi-agent coordination.
\newblock In {\em Proceedings of the 2005, American Control Conference, 2005.},
  pages 1859--1864 vol. 3, 2005.

\bibitem{fell2011role}
Juergen Fell and Nikolai Axmacher.
\newblock The role of phase synchronization in memory processes.
\newblock {\em Nature reviews neuroscience}, 12(2):105--118, 2011.

\bibitem{varela2001brainweb}
Francisco Varela, Jean-Philippe Lachaux, Eugenio Rodriguez, and Jacques
  Martinerie.
\newblock The brainweb: phase synchronization and large-scale integration.
\newblock {\em Nature reviews neuroscience}, 2(4):229--239, 2001.

\bibitem{buzsaki2004neuronal}
Gyorgy Buzsaki and Andreas Draguhn.
\newblock Neuronal oscillations in cortical networks.
\newblock {\em Science}, 304(5679):1926--1929, 2004.

\bibitem{friedkin2016network}
Noah~E Friedkin, Anton~V Proskurnikov, Roberto Tempo, and Sergey~E Parsegov.
\newblock Network science on belief system dynamics under logic constraints.
\newblock {\em Science}, 354(6310):321--326, 2016.

\bibitem{Ye2020ContinuoustimeOD}
Mengbin Ye, Minh~Hoang Trinh, Young-Hun Lim, Brian. D.~O. Anderson, and
  Hyo-Sung Ahn.
\newblock Continuous-time opinion dynamics on multiple interdependent topics.
\newblock {\em Automatica}, 115:108884, 2020.

\bibitem{wensink2012meso}
Henricus~H Wensink, J{\"o}rn Dunkel, Sebastian Heidenreich, Knut Drescher,
  Raymond~E Goldstein, Hartmut L{\"o}wen, and Julia~M Yeomans.
\newblock Meso-scale turbulence in living fluids.
\newblock {\em Proceedings of the national academy of sciences},
  109(36):14308--14313, 2012.

\bibitem{dunkel2013fluid}
J{\"o}rn Dunkel, Sebastian Heidenreich, Knut Drescher, Henricus~H Wensink,
  Markus B{\"a}r, and Raymond~E Goldstein.
\newblock Fluid dynamics of bacterial turbulence.
\newblock {\em Physical review letters}, 110(22):228102, 2013.

\bibitem{tuna2019synchronization}
S~Emre Tuna.
\newblock Synchronization of small oscillations.
\newblock {\em Automatica}, 107:154--161, 2019.

\bibitem{zhao2016localizability}
Shiyu Zhao and Daniel Zelazo.
\newblock Localizability and distributed protocols for bearing-based network
  localization in arbitrary dimensions.
\newblock {\em Automatica}, 69:334--341, 2016.

\bibitem{barooah2008estimation}
Prabir Barooah and Joao~P Hespanha.
\newblock Estimation from relative measurements: Electrical analogy and large
  graphs.
\newblock {\em IEEE Transactions on Signal Processing}, 56(6):2181--2193, 2008.

\bibitem{TRINH2018415}
Minh~Hoang Trinh, Chuong Van Nguyen, Young-Hun Lim, and Hyo-Sung Ahn.
\newblock Matrix-weighted consensus and its applications.
\newblock {\em Automatica}, 89:415 -- 419, 2018.

\bibitem{Pan2019}
L.~Pan, H.~Shao, M.~Mesbahi, Y.~Xi, and D.~Li.
\newblock Bipartite consensus on matrix-valued weighted networks.
\newblock {\em IEEE Transactions on Circuits and Systems II: Express Briefs},
  66(8):1441--1445, 2019.

\bibitem{wang2022characterizing}
Chongzhi Wang, Lulu Pan, Haibin Shao, Dewei Li, and Yugeng Xi.
\newblock Characterizing bipartite consensus on signed matrix-weighted networks
  via balancing set.
\newblock {\em Automatica}, 141:110237, 2022.

\bibitem{Partridge1982}
Brian~L. Partridge.
\newblock The structure and function of fish schools.
\newblock {\em Scientific American}, 246(6):114--123, 1982.

\bibitem{Parrish2002}
Julia~K. Parrish, Steven~V. Viscido, and Daniel Grünbaum.
\newblock Self-organized fish schools: An examination of emergent properties.
\newblock {\em Biological Bulletin}, 202(3):296--305, 2002.

\bibitem{honey2007network}
Christopher~J Honey, Rolf K{\"o}tter, Michael Breakspear, and Olaf Sporns.
\newblock Network structure of cerebral cortex shapes functional connectivity
  on multiple time scales.
\newblock {\em Proceedings of the National Academy of Sciences},
  104(24):10240--10245, 2007.

\bibitem{isaacson2011inhibition}
Jeffry~S Isaacson and Massimo Scanziani.
\newblock How inhibition shapes cortical activity.
\newblock {\em Neuron}, 72(2):231--243, 2011.

\bibitem{Sadilek2015}
Maximilian Sadilek and Stefan Thurner.
\newblock Physiologically motivated multiplex kuramoto model describes phase
  diagram of cortical activity.
\newblock {\em Scientific Reports}, 5(1):10015, 2015.

\bibitem{Purves2012}
Dale Purves, George~J. Augustine, David Fitzpatrick, William~C. Hall,
  Anthony-Samuel LaManita, and Leonard~E. White.
\newblock Neuroscience, 5th ed., 2012.

\bibitem{kuramoto2003chemical}
Yoshiki Kuramoto.
\newblock {\em Chemical oscillations, waves, and turbulence}.
\newblock Courier Corporation, 2003.

\bibitem{strogatz2000kuramoto}
Steven~H Strogatz.
\newblock From kuramoto to crawford: exploring the onset of synchronization in
  populations of coupled oscillators.
\newblock {\em Physica D: Nonlinear Phenomena}, 143(1-4):1--20, 2000.

\bibitem{van1993lyapunov}
JL~Van~Hemmen and WF~Wreszinski.
\newblock Lyapunov function for the kuramoto model of nonlinearly coupled
  oscillators.
\newblock {\em Journal of Statistical Physics}, 72(1):145--166, 1993.

\bibitem{buck1968mechanism}
John Buck and Elisabeth Buck.
\newblock Mechanism of rhythmic synchronous flashing of fireflies: Fireflies of
  southeast asia may use anticipatory time-measuring in synchronizing their
  flashing.
\newblock {\em Science}, 159(3821):1319--1327, 1968.

\bibitem{acebron2005kuramoto}
Juan~A Acebron, Luis~L Bonilla, Conrad J~Perez Vicente, Felix Ritort, and
  Renato Spigler.
\newblock The kuramoto model: A simple paradigm for synchronization phenomena.
\newblock {\em Reviews of modern physics}, 77(1):137, 2005.

\bibitem{childs2008stability}
Lauren~M Childs and Steven~H Strogatz.
\newblock Stability diagram for the forced kuramoto model.
\newblock {\em Chaos: An Interdisciplinary Journal of Nonlinear Science},
  18(4):043128, 2008.

\bibitem{antonsen2008external}
TM~Antonsen~Jr, RT~Faghih, M~Girvan, E~Ott, and J~Platig.
\newblock External periodic driving of large systems of globally coupled phase
  oscillators.
\newblock {\em Chaos: An Interdisciplinary Journal of Nonlinear Science},
  18(3):037112, 2008.

\bibitem{hoppensteadt1997weakly}
Frank~C Hoppensteadt and Eugene~M Izhikevich.
\newblock {\em Weakly connected neural networks}, volume 126.
\newblock Springer Science \& Business Media, 1997.

\bibitem{tanaka2014solvable}
Takuma Tanaka.
\newblock Solvable model of the collective motion of heterogeneous particles
  interacting on a sphere.
\newblock {\em New Journal of Physics}, 16(2):023016, 2014.

\bibitem{zhu2013synchronization}
Jiandong Zhu.
\newblock Synchronization of kuramoto model in a high-dimensional linear space.
\newblock {\em Physics Letters A}, 377(41):2939--2943, 2013.

\bibitem{chandra2019continuous}
Sarthak Chandra, Michelle Girvan, and Edward Ott.
\newblock Continuous versus discontinuous transitions in the d-dimensional
  generalized kuramoto model: Odd d is different.
\newblock {\em Physical Review X}, 9(1):011002, 2019.

\bibitem{zhang2015explosive}
Xiyun Zhang, Stefano Boccaletti, Shuguang Guan, and Zonghua Liu.
\newblock Explosive synchronization in adaptive and multilayer networks.
\newblock {\em Physical review letters}, 114(3):038701, 2015.

\bibitem{watanabe1994constants}
Shinya Watanabe and Steven~H Strogatz.
\newblock Constants of motion for superconducting josephson arrays.
\newblock {\em Physica D: Nonlinear Phenomena}, 74(3-4):197--253, 1994.

\bibitem{dorfler2012synchronization}
Florian Dorfler and Francesco Bullo.
\newblock Synchronization and transient stability in power networks and
  nonuniform kuramoto oscillators.
\newblock {\em SIAM Journal on Control and Optimization}, 50(3):1616--1642,
  2012.

\bibitem{dorfler2013synchronization}
Florian D{\"o}rfler, Michael Chertkov, and Francesco Bullo.
\newblock Synchronization in complex oscillator networks and smart grids.
\newblock {\em Proceedings of the National Academy of Sciences},
  110(6):2005--2010, 2013.

\bibitem{Millan2019ExplosiveHK}
Ana~P. Mill'an, Joaquin~J. Torres, and Ginestra Bianconi.
\newblock Explosive higher-order kuramoto dynamics on simplicial complexes.
\newblock {\em Physical review letters}, 124 21:218301, 2019.

\bibitem{Skardal2019AbruptDA}
Per~Sebastian Skardal and Alex Arenas.
\newblock Abrupt desynchronization and extensive multistability in globally
  coupled oscillator simplexes.
\newblock {\em Physical review letters}, 122 24:248301, 2019.

\bibitem{Skardal2020HigherOI}
Per~Sebastian Skardal and Alex Arenas.
\newblock Higher order interactions in complex networks of phase oscillators
  promote abrupt synchronization switching.
\newblock {\em Communications Physics}, 3, 2020.

\bibitem{Skardal2022MultistabilityIC}
Per~Sebastian Skardal, Sabina Adhikari, and Juan~G. Restrepo.
\newblock Multistability in coupled oscillator systems with higher-order
  interactions and community structure.
\newblock {\em Chaos}, 33 2:023140, 2022.

\bibitem{Reimann2017}
Michael~W. Reimann, Max Nolte, Martina Scolamiero, Katharine Turner, Rodrigo
  Perin, Giuseppe Chindemi, Pawel Dlotko, Ran Levi, Kathryn Hess, and Henry
  Markram.
\newblock Cliques of neurons bound into cavities provide a missing link between
  structure and function.
\newblock {\em Frontiers in Computational Neuroscience}, 11, 2017.

\bibitem{Giusti2016TwosCT}
Chad Giusti, Robert Ghrist, and Danielle~S. Bassett.
\newblock Two's a company, three (or more) is a simplex.
\newblock {\em Journal of Computational Neuroscience}, 41:1 -- 14, 2016.

\bibitem{Sizemore2016CliquesAC}
Ann~E. Sizemore, Chad Giusti, Ari~E. Kahn, Jean~M. Vettel, Richard~F. Betzel,
  and Danielle~S. Bassett.
\newblock Cliques and cavities in the human connectome.
\newblock {\em Journal of Computational Neuroscience}, 44:115 -- 145, 2016.

\bibitem{Bick2016}
Christian Bick and Ana Rodrigues.
\newblock Chaos in generically coupled phase oscillator networks with
  nonpairwise interactions.
\newblock {\em Chaos: An Interdisciplinary Journal of Nonlinear Science}, 26,
  05 2016.

\bibitem{Dai2020DiscontinuousTA}
Xiangfeng Dai, X.~Li, H.~Guo, D.~Jia, M.~Perc, Pouya Manshour, Z.~Wang, and
  Stefano Boccaletti.
\newblock Discontinuous transitions and rhythmic states in the d-dimensional
  kuramoto model induced by a positive feedback with the global order
  parameter.
\newblock {\em Physical review letters}, 125 19:194101, 2020.

\bibitem{Wiley2006TheSO}
Daniel~A Wiley, Steven~H. Strogatz, and Michelle Girvan.
\newblock The size of the sync basin.
\newblock {\em Chaos}, 16 1:015103, 2006.

\bibitem{Delabays2017TheSO}
Robin Delabays, Melvyn Tyloo, and Philippe Jacquod.
\newblock The size of the sync basin revisited.
\newblock {\em Chaos}, 27 10:103109, 2017.

\bibitem{Ostrowski1959berEV}
Alexander Ostrowski.
\newblock {\"U}ber eigenwerte von produkten hermitescher matrizen.
\newblock {\em Abhandlungen aus dem Mathematischen Seminar der Universit{\"a}t
  Hamburg}, 23:60--68, 1959.

\bibitem{li2008synchronization}
Ping Li and Zhang Yi.
\newblock Synchronization of kuramoto oscillators in random complex networks.
\newblock {\em Physica A: Statistical Mechanics and its Applications},
  387(7):1669--1674, 2008.

\bibitem{patra2016statistics}
Soumen~K Patra and Anandamohan Ghosh.
\newblock Statistics of lyapunov exponent spectrum in randomly coupled kuramoto
  oscillators.
\newblock {\em Physical Review E}, 93(3):032208, 2016.

\bibitem{Coss2018}
Owen Coss, Jonathan~D. Hauenstein, Hoon Hong, and Daniel~K. Molzahn.
\newblock Locating and counting equilibria of the kuramoto model with rank-one
  coupling.
\newblock {\em SIAM Journal on Applied Algebra and Geometry}, 2(1):45--71,
  2018.

\bibitem{lameu2018alterations}
Ewandson~Luiz Lameu, Elbert~EN Macau, FS~Borges, Kelly~Cristiane Iarosz,
  Iber{\^e}~Luiz Caldas, Rafael~Ribaski Borges, PR~Protachevicz, Ricardo~Luiz
  Viana, and Antonio~Marcos Batista.
\newblock Alterations in brain connectivity due to plasticity and synaptic
  delay.
\newblock {\em The European Physical Journal Special Topics}, 227:673--682,
  2018.

\bibitem{Buzanello2022}
Guilhermo~L. Buzanello, Ana Elisa~D. Barioni, and Marcus A.~M. de~Aguiar.
\newblock {Matrix coupling and generalized frustration in Kuramoto
  oscillators}.
\newblock {\em Chaos: An Interdisciplinary Journal of Nonlinear Science},
  32(9), 09 2022.
\newblock 093130.

\bibitem{gomez2011explosive}
Jes{\'u}s G{\'o}mez-Gardenes, Sergio G{\'o}mez, Alex Arenas, and Yamir Moreno.
\newblock Explosive synchronization transitions in scale-free networks.
\newblock {\em Physical review letters}, 106(12):128701, 2011.

\bibitem{andrew1989seizure}
R~David Andrew, Mitchell Fagan, Barbara~A Ballyk, and Andrei~S Rosen.
\newblock Seizure susceptibility and the osmotic state.
\newblock {\em Brain research}, 498(1):175--180, 1989.

\bibitem{wang2017small}
Zhenhua Wang, Changhai Tian, Mukesh Dhamala, and Zonghua Liu.
\newblock A small change in neuronal network topology can induce explosive
  synchronization transition and activity propagation in the entire network.
\newblock {\em Scientific reports}, 7(1):561, 2017.

\bibitem{wang2013brain}
Megan Wang, Daniel Arteaga, and Biyu~J He.
\newblock Brain mechanisms for simple perception and bistable perception.
\newblock {\em Proceedings of the National Academy of Sciences},
  110(35):E3350--E3359, 2013.

\bibitem{Ott2008}
Edward Ott and Thomas~M. Antonsen.
\newblock {Low dimensional behavior of large systems of globally coupled
  oscillators}.
\newblock {\em Chaos: An Interdisciplinary Journal of Nonlinear Science},
  18(3), 09 2008.
\newblock 037113.

\bibitem{ott2009long}
Edward Ott and Thomas~M Antonsen.
\newblock Long time evolution of phase oscillator systems.
\newblock {\em Chaos: An interdisciplinary journal of nonlinear science},
  19(2):023117, 2009.

\bibitem{chandra2019complexity}
Sarthak Chandra, Michelle Girvan, and Edward Ott.
\newblock Complexity reduction ansatz for systems of interacting orientable
  agents: Beyond the kuramoto model.
\newblock {\em Chaos: An Interdisciplinary Journal of Nonlinear Science},
  29(5):053107, 2019.

\bibitem{Chandra2019ObservingMT}
Sarthak Chandra and Edward Ott.
\newblock Observing microscopic transitions from macroscopic bursts:
  Instability-mediated resetting in the incoherent regime of the d-dimensional
  generalized kuramoto model.
\newblock {\em Chaos}, 29 3:033124, 2019.

\bibitem{Pecora2014}
Louis~M. Pecora, Francesco Sorrentino, Aaron~M. Hagerstrom, Thomas~E. Murphy,
  and Rajarshi Roy.
\newblock Cluster synchronization and isolated desynchronization in complex
  networks with symmetries.
\newblock {\em Nature Communications}, 5(1):4079, 2014.

\bibitem{lodi2021one}
Matteo Lodi, Francesco Sorrentino, and Marco Storace.
\newblock One-way dependent clusters and stability of cluster synchronization
  in directed networks.
\newblock {\em Nature Communications}, 12(1):4073, 2021.

\bibitem{Cho2017}
Young~Sul Cho, Takashi Nishikawa, and Adilson~E. Motter.
\newblock Stable chimeras and independently synchronizable clusters.
\newblock {\em Phys. Rev. Lett.}, 119:084101, Aug 2017.

\bibitem{mukhametov1977interhemispheric}
LM~Mukhametov, A~Ya Supin, and IG~Polyakova.
\newblock Interhemispheric asymmetry of the electroencephalographic sleep
  patterns in dolphins.
\newblock {\em Brain research}, 1977.

\bibitem{Rattenborg2000BehavioralNA}
Niels~C. Rattenborg, Charles~J. Amlaner, and Steven~L. Lima.
\newblock Behavioral, neurophysiological and evolutionary perspectives on
  unihemispheric sleep.
\newblock {\em Neuroscience \& Biobehavioral Reviews}, 24:817--842, 2000.

\bibitem{Raychowdhury2019}
Arijit Raychowdhury, Abhinav Parihar, Gus~Henry Smith, Vijaykrishnan Narayanan,
  Gyorgy Csaba, Matthew Jerry, Wolfgang Porod, and Suman Datta.
\newblock Computing with networks of oscillatory dynamical systems.
\newblock {\em Proceedings of the IEEE}, 107(1):73--89, 2019.

\bibitem{Csaba2020}
Gyorgy Csaba and Wolfgang Porod.
\newblock {Coupled oscillators for computing: A review and perspective}.
\newblock {\em Applied Physics Reviews}, 7(1), 01 2020.
\newblock 011302.

\bibitem{Hoppensteadt2000}
F.C. Hoppensteadt and E.M. Izhikevich.
\newblock Pattern recognition via synchronization in phase-locked loop neural
  networks.
\newblock {\em IEEE Transactions on Neural Networks}, 11(3):734--738, 2000.

\bibitem{Strogatz1991StabilityOI}
Steven~H. Strogatz and Renato Mirollo.
\newblock Stability of incoherence in a population of coupled oscillators.
\newblock {\em Journal of Statistical Physics}, 63:613--635, 1991.

\end{thebibliography}

\end{document}